\documentclass[referee]{chjaa}
\usepackage{graphicx,times}
\usepackage{deluxetable}
\input{epsf.sty}
\input{psfig.sty}
\newcommand{\teff} {\mbox{$T_{\rm eff}$}}
\newcommand{\logg} {\mbox{{\rm log}$g$}}
\newcommand{\feh}  {\mbox{\rm [Fe/H]}}

\newcommand{\nafe} {\mbox{\rm [Na/Fe]}}
\newcommand{\mgfe} {\mbox{\rm [Mg/Fe]}}
\newcommand{\alfe} {\mbox{\rm [Al/Fe]}}
\newcommand{\sife} {\mbox{\rm [Si/Fe]}}
\newcommand{\cafe} {\mbox{\rm [Ca/Fe]}}
\newcommand{\scfe} {\mbox{\rm [Sc/Fe]}}
\newcommand{\tife} {\mbox{\rm [Ti/Fe]}}
\newcommand{\vfe}  {\mbox{\rm [V/Fe]}}
\newcommand{\crfe} {\mbox{\rm [Cr/Fe]}}
\newcommand{\mnfe} {\mbox{\rm [Mn/Fe}}
\newcommand{\nife} {\mbox{\rm [Ni/Fe]}}
\newcommand{\yfe}  {\mbox{\rm [Y/Fe]}}
\newcommand{\zrfe} {\mbox{\rm [Zr/Fe]}}
\newcommand{\lafe} {\mbox{\rm [La/Fe]}}
\newcommand{\eufe} {\mbox{\rm [Eu/Fe]}}
\newcommand{\bafe} {\mbox{\rm [Ba/Fe]}}
\begin{document}
\title {Abundance Analysis of Barium Stars}

\volnopage{Vol.0 (200x) No.0, 000--000}
\setcounter{page}{1}
\author{G. Q. Liu\inst{1,2}\mailto{}
        \and Y. C. Liang\inst{1}\mailto{}
    \and L. Deng\inst{1}
          }

\institute{National Astronomical Observatories, Chinese Academy of
Sciences, Beijing 100012, P. R. China\\
       \email{lgq@bao.ac.cn, ycliang@bao.ac.cn}
\and
Graduate University of Chinese Academy of Sciences, Beijing
100049, P. R. China}

   \date{Received; accepted}

  \abstract
   {We obtain the chemical abundances of six barium stars and two CH subgiant stars based
on the high signal-to-noise ratio and high resolution Echelle
spectra. The neutron capture process elements Y, Zr, Ba, La, Eu
show obvious overabundance relative to the Sun, for example, their
[Ba/Fe] values are from 0.45 to 1.27. Other elements, including
Na, Mg, Al, Si, Ca, Sc, Ti, V, Cr, Mn, Ni, show comparable
abundances to the Solar ones, and their \feh~cover a range from
$-$0.40 to 0.21, which means they belong to Galactic disk. The
predicts of the theoretical model of wind accretion for binary
systems can explain the observed abundance patterns of the neutron
capture process elements in these stars, which means that their
overabundant heavy-elements could be caused by accreting the
ejecta of AGB stars, the progenitors of the present white dwarf
companions in the binary systems.
               }
\authorrunning{G. Q. Liu, Y. C. Liang \& L. Deng}
\titlerunning{Abundance Analysis of Barium Stars}
   \maketitle
%
%
\section{Introduction}
As first identified by Bidelman \& Keeman (\cite{bk51}), barium
stars appear as a distinct group of chemically peculiar red
giants. These G and K giants show enhanced features of
\ion{Ba}{ii}, \ion{Sr}{ii}, CH, CN and sometimes C$_{2}$ lines.
The following studies also found enhanced abundances of some other
heavy elements, e.g. Y, Zr, La, Ce, Pr, Nd and Sm.

Since Burbidge et al. (\cite{bbfh57}) suggested the elements
heavier than iron are synthesized in the interior of asymptotic
giant branch (AGB) stars through the slow neutron capture process
(s-process) (the rapid neutron capture process, r-process, occurs
in supernova explosion), one generally believe that the
overabundant heavy elements of Ba stars could be caused by binary
accretion because they should not be evolved to the thermal pulse
(TP) AGB stage to synthesize these heavy elements due to their low
luminosity and the absence of the unstable nucleus $^{99}$Tc
($\tau_{1\over 2}$=2$\times 10^{5}$yr) (see Liang et al. 2000,
2003 and references therein). The binarity and heavy-element
abundances of Ba stars have been studied by many researchers
(Griffin \cite{gr80}; Jorissen \& Mayor \cite{jm88}; McClure et
al. \cite{mc80}; McClure \cite{mc83}; McClure \& Woodsworth
\cite{mw90}; Jorissen et al. 1998; Liang et al. 2000, 2003; Liu et
al. 2000; L\"{u} et al. 1991; Han et al. 1995; Za$\check{c}$s
1994; Smiljanic et al. 2007). These Ba stars could have accreted
the matter ejected by their companions (the former AGB stars, the
present white dwarfs) about 1$\times 10^{6}$ years ago through
wind accretion, disk accretion or common envelope ejection (Han et
al. 1995; Jorissen et al. 1998; Liang et al. 2000).

\begin{table}[]
\tablenum{1}
\caption[]{Basic data of the sample stars.
The HD identifications, spectral types, $B-V$ color,
trigonometric parallaxes and their errors.}
\begin{center}\begin{tabular}{llllrr} \hline\hline
\noalign{\smallskip}
\noalign{\smallskip}
HD& Sp.&V$_{\rm mag}$ & $B-V$ &  $\pi$ & $\sigma_\pi$ \\
      &      &&         &             (mas) &
           \\\hline
\noalign{\smallskip}
4395  & G5   &7.70& 0.69  &  9.16 &  1.12 \\
180622& K2   &7.63& 1.25  &  3.37 &  1.04 \\
201657& K2   &8.00& 1.27  &  4.49 &  1.07 \\
201824& K0   &8.90& 1.09  &  0.56 &  1.56 \\
210946& K0   &8.08& 1.095 &  3.42 &  1.14 \\
211594& K0   &8.05& 1.143 &  4.59 &  1.18 \\
216219& G0IIp&7.44& 0.64  &  10.74&  0.93 \\
223617& G5   &6.91& 1.155 &  4.61 &  0.95 \\
\noalign{\smallskip}
\hline
\end{tabular}
\end{center}
\end{table}

At present, a large sample of Ba stars have been measured their
binary orbital elements (Carquillat et al. \cite{cjug98}; Udry et
al. \cite{ujmv98}, \cite{umvjpgl98}; Jorissen et al. 1998),
absolute magnitudes and kinematics (G\'{o}mez et al. \cite{glg97};
Mennessier et al. \cite{mlf97}). However, the corresponding
heavy-element abundances have not been obtained from high
resolution observations, which is very useful to understand the
formation scenario of Ba stars, but need lots of telescope time
and lots of efforts on data analysis. Therefore, we propose to
observe the high resolution and high signal-to-noise (S/N) ratio
spectra of a sample of Ba stars to obtain their chemical
abundances, hence, to understand their formation scenario by
combining with their binary orbital elements. Moreover, by taking
advantage of the present high precision Hipparcos data, precise
photometric parameters, improved methods to determine stellar
atmospheric parameters and developed stellar evolutionary tracks
etc., the reliable chemical abundances of stars could be obtained
from the spectra. We could also understand the formation scenario
of Ba stars from theoretical models by comparing the model
predicts with the observed abundances, e.g. the angular momentum
conservation model of wind accretion of Ba binaries (Liang et al.
2000; Liu et al. 2000; Boffin \& Jorissen \cite{bj88}).

 This paper is organized
as follows. Description of the spectral observations and data
reduction for the sample stars are presented in Section 2. In
Section 3 the derived stellar atmospheric parameters are
presented. Stellar atmosphere model, spectral lines and their
measured equivalent widths (EWs) are described in Section 4. The
analysis on abundance results is given in Section 5. The predicted
abundances from wind accretion model are presented and compared
with the observed abundance patterns in Section 6. The discussions
and conclusions are given in Section 7.

%
%
\section{Observations and data reduction}
The sample stars have been firstly identified as mild or strong Ba
stars in L\"{u} et al. (1991) and have been obtained their binary
orbital elements (e.g. orbital period and eccentricity) in
Jorissen et al. (1998) except HD\,4395. HD\,4395 and
HD\,216219 were classified as CH subgiants (Luck \& Bond 1982;
Sneden 1983; Krishnaswamy \& Sneden 1985; Lambert et al. 1993;
Smith et al. 1993; Preston \& Sneden 2001), but HD\,216219 has
also been classified as mild Ba star by  L\"{u} et al. (1991) and
Jorissen et al. (1998). A common point of view is that CH
subgiants also belong to binary systems and their overabundances
of s-process elements are caused by accreting the ejected material
of the companion AGB progenitors, which is the same scenario as
the Ba stars. The CH subgiants could evolve to be the classical Ba
stars (Luck \& Bond 1982; Smith et al. 1993). In this work we take
these two stars as the same with other samples to study their
abundances and formation scenario. These two common stars with
Smith et al. (1993) are good examples to compare our EW
measurements, atmospheric parameters and abundances with theirs
carefully.

Table~1 presents the basic parameters of the sample stars. The
Column (1)-(6) consequently present their HD identifications,
spectral type and luminosity class, visual magnitudes, $B-V$ color
index, trigonometric parallaxes and the corresponding errors taken
from SIMBAD database.

\begin{figure}
\vspace{2mm}
\hspace{3mm}\psfig{figure=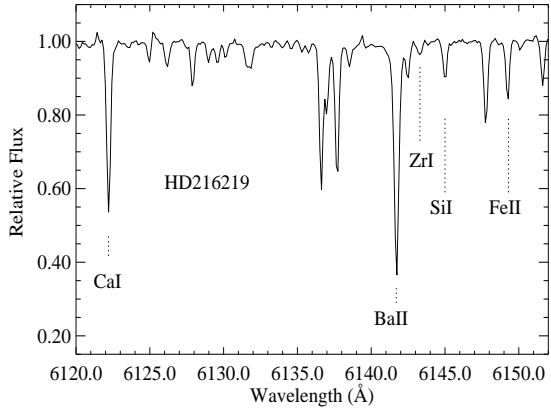,width=80mm,height=60mm,angle=0.0}
\caption{A portion of spectrum of the sample star HD 216219 in the
wavelength range 6120{\AA}$-$6152{\AA}.
Major features include \ion{Ca}{i} 6122.226 {\AA}, \ion{Ba}{ii} 6141.727 {\AA},
\ion{Zr}{i} 6143.180 {\AA}, \ion{Si}{i} 6145.020 {\AA} and \ion{Fe}{ii} 6149.249 {\AA}.}
\label{portion}
\end{figure}

The spectroscopic observations were carried out with the Coud\'{e}
Echelle Spectrograph of National Astronomical Observatories (NAOC)
mounted on the 2.16 m telescope at Xinglong station (Xinglong, P.
R. China). The detector was a Tektronix CCD with $ 1024\times
1024$ pixels ($24\times 24~\mu m^{2}$ each in size). The
wavelength coverage of total spectra is roughly from 5500$-$8000
\AA\ over 34 orders. The spectra were observed during September
12$-$14, 2000 and most of them had S/N $>$ 100. Figure
\ref{portion} presents the spectrum of HD 216219 showing main
features of absorption in the region around \ion{Ba}{ii}
$\lambda$6141.727 line.

Data reductions of all the spectra were carried out through ECHELLE package
in the MIDAS environment by standard routines proceeding with order identification,
background subtraction, flat-field correction, order extraction, wavelength
calibration with a thorium-argon lamp calibration frame.
Bias, dark current and scattered light corrections were taken into account
in the background subtraction.
The pixel-to-pixel sensitivity variations were corrected
by using the flat-field.
The EWs of the spectral lines are measured from the
normalized spectra corrected by radial velocity, which were
measured from more than 20 absorption lines at least.
The selected spectral lines for abundance analysis are unblended or slightly blended
and have reliable atomic data.

\begin{figure}
\centering \hspace{3mm}\psfig{figure=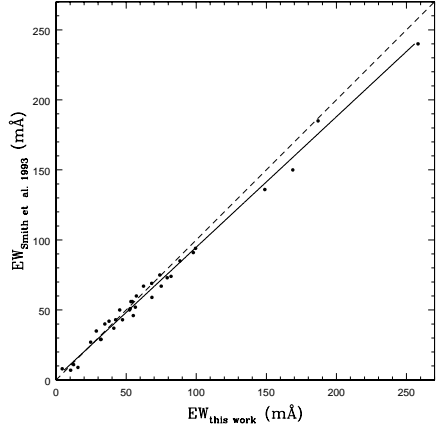,width=60mm}
\caption{Comparison of equivalent width measurements of 35
common absorption lines in HD 216219 and HD 4395 in this work and
Smith et al. (\cite{scl93}). The solid line is the least square
fit to the points (Eq.(\ref{eqEWfit})), and the dashed line refers
to the one-to-one relation.} \label{comp}
\end{figure}

The EWs of spectral line were measured by applying two different
methods: direct integration of the line profile and Gaussian
fitting. The latter is preferable in the case of faint lines (EW
$<$ 20 m\AA), but unsuitable for the strong lines in which the
damping wings contribute significantly to the equivalent width.
The final EWs are weighted averages of these two measurements,
depending on the line intensity (see Zhao et al. \cite{zqcl00} for
details). The EW values of 110$-$180 lines
 in the wavelength range from 5500$-$8100
\AA\ were obtained for each of the sample stars.
 Table 2 presents the final EWs of
the lines measured in the spectra of the sample stars as input
data for the abundance analysis. Since the very weak lines would
lead to an increase of random errors in the abundance
determination and the too strong lines are not so sensitive to
abundance, we use the lines with EWs=$10-200$
m\AA~ for abundance analysis, and most of them within $20-150$~m\AA~
except some of the s-process elements. The reliability of our EW
measurements have been confirmed by the consistence in the
comparison between our data and those of Smith et
al.~(\cite{scl93}) for 35 common lines of HD 216219 and HD 4395.
This comparison is shown in Figure~\ref{comp}. The systematic
difference between the two sets of data is small and could be
given by a linear least square fit as:
\begin{equation}
EW_{\rm Smith93} = 0.92(\pm 0.02)EW_{\rm this~work}+2.19(\pm 1.30) m\AA,
\label{eqEWfit}
\end{equation}
with the standard deviation of 0.061.

\begin{figure*}
\centering
\hspace{3mm}\psfig{figure=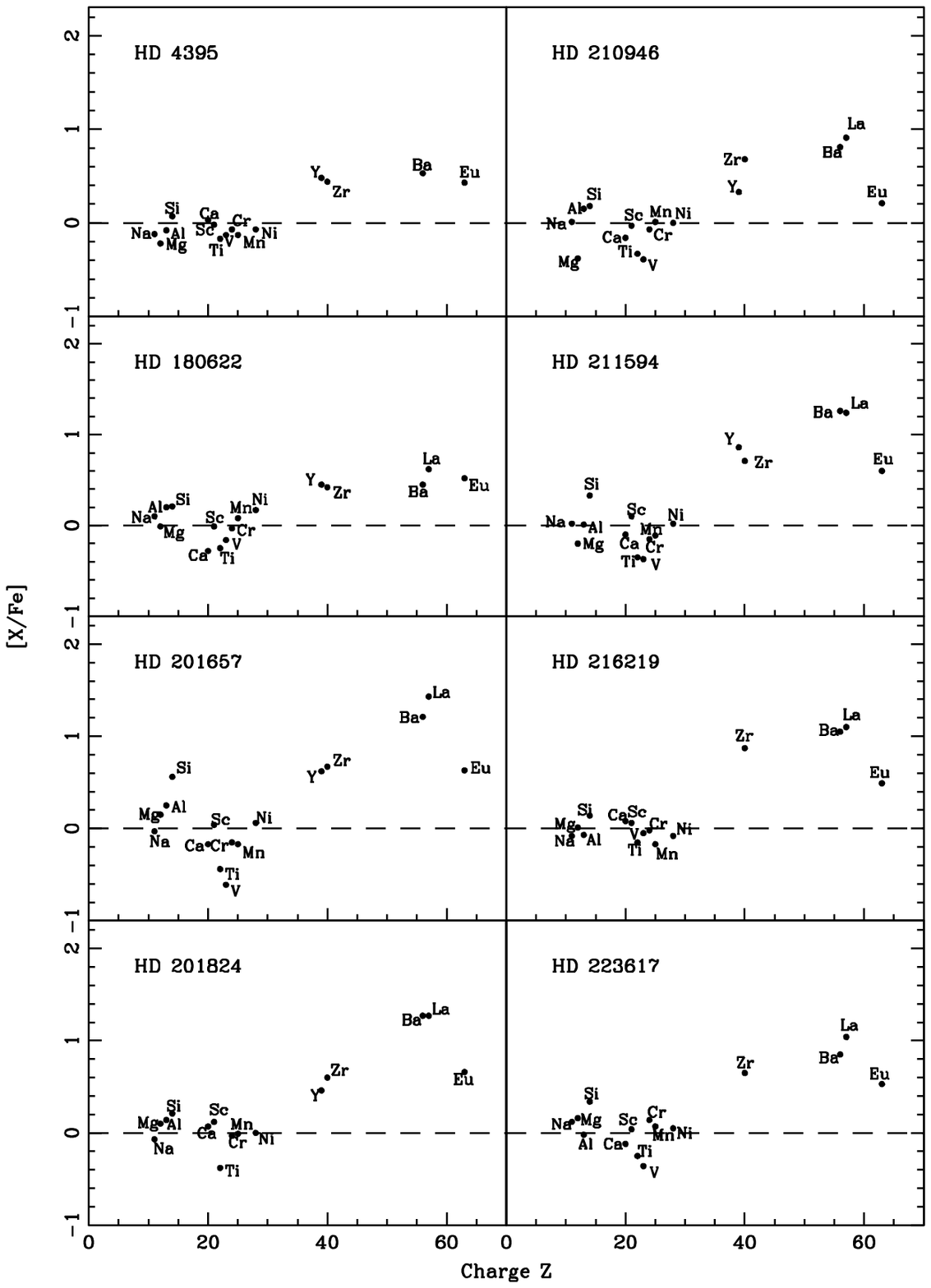,width=120mm}
\caption{The abundance patterns of sample stars.}
\label{abun}
\end{figure*}

{
\begin{table*}
\tablenum{3}
{\begin{center}
\caption[] {Atmospheric
parameters of the sample stars: effective temperature (\teff),
surface gravity (log\,$g$), mass (M/M$_{\odot}$) and their
uncertainties, microturbulence velocity ($\xi_t$) and metallicity
(\feh).}
\begin{tabular}{lrrrrr}
\hline\hline
\noalign{\smallskip}
HD &\teff & log\,$g$& M/M$_{\odot}$& $\xi_t$ & [Fe/H]  \\
\noalign{\smallskip}
\hline
\noalign{\smallskip}
4395  & 5447&3.60 & 1.60(+0.13,$-0.10$) & 1.3&$-$0.16 \\
180622& 4391&2.24 & 1.92(+1.07,$-0.36$) &1.5 &0.21    \\
201657& 4284&2.17 & 0.78(+0.08,$-0.02$) & 1.7&$-$0.31 \\
201824& 4552&1.67 & 4.58($-$,$-3.32$) & 1.5&$-$0.40 \\
210946& 4577&2.42 & 1.40(+0.72,$-0.28$) & 1.6&$-$0.22 \\
211594& 4490&2.44 & 0.90(+0.41,$-0.09$) & 1.6&$-$0.23 \\
216219& 5553&3.64 & 1.48(+0.07,$-0.08$) & 1.4&$-$0.34 \\
223617& 4501&2.27 & 1.78(+0.47,$-0.33$) & 1.5&$-$0.10 \\
\noalign{\smallskip} \hline\hline
\end{tabular}
\end{center}
}
\end{table*}
}

\section{Stellar atmospheric parameters}

The stellar atmospheric model is composed of four basic
parameters, i.e., effective temperature, surface gravity,
metallicity and microturbulent velocity. In this section, we
describe the determinations of these four model-atmosphere
parameters.

\subsection{Effective temperature}

Effective temperatures $T_{\rm eff}$ were determined from \feh~
and $\emph{B-V}$ color indices by using the empirical calibration
of Alonso et al. (\cite{aam99} \& \cite{aam01}), which is suitable
for giant stars. We use $\emph{B-V}$ color here since the
$\emph{B-V}$ data are more complete than other color indices for
the sample stars. Considering the uncertainties of photometric
data, [Fe/H], and the errors in the calibration relation, we
estimate the uncertainty in \teff ~is about 100\,K for our sample
stars.

\subsection{Surface Gravity}

Based on
the $\emph{Hipparcos}$ parallaxes, precise value of the surface
gravity of nearby stars can be obtained using the following
relations:
\begin{eqnarray}
 \log \frac{g}{g_{\odot}}&=&\log \frac{\mathcal M}{{\mathcal M}_{\odot}}
+4 \log \frac{\teff}{\teff_{\odot}}+
0.4 \left(M_{\mathrm{bol}}-M_{\mathrm{bol},\odot}\right)  \label{eq:logg}
\end{eqnarray}
and
\begin{eqnarray}
 M_{\mathrm{bol}}&=& V+BC+5 \log \pi + 5,\label{eq:Mv}
\end{eqnarray}
where ${\mathcal M}$ is the stellar mass, $M_{\mathrm{bol}}$ the
absolute bolometric magnitude, $V$ the visual magnitude, $BC$ the
bolometric correction, and $\pi$ the parallax. We adopt solar
value log\,$g_\odot$=4.44, $T_{{\rm eff},\odot}$=5770 K, $M_{{\rm
bol},\odot}$=4.77 mag. The parallax $\pi$ and its errors are taken
from the Hipparcos Satellite observations (ESA 1997). Stellar mass
was determined from the position of the star in $M_{\rm
bol}-$log\,$T_{\rm eff}$ diagram. We adopt the stellar
evolutionary tracks of Yonsei-Yale (Yi et al. \cite{ykd03}), whose
isochrones determined with high quality observational data cover
the stage from pre-main-sequence birthline to the helium-core
flash. The uncertainty of log$g$ estimated by this method is about
0.2 dex generally for our sample stars.

%
\subsection{Metal abundance}
The initial metallicities of the sample stars in their model
atmospheres were taken from literature if available. Otherwise, we
adopt \feh=0 as the initial value and then the final model
metallicitiy derived from the consistency with the other
parameters in the abundance calculation. The estimated error in
[Fe/H] is about 0.1 dex.
\begin{table}
\tablenum{4}
\caption{Element abundances of the sample stars.}
{\tiny
   $$
\setlength\tabcolsep{3pt}
\begin{tabular}{lrrrrrrrrrrrrrrrr}
\hline\hline
\noalign{\smallskip}
&& \multicolumn{3}{c}{HD 4395} && \multicolumn{3}{c}{HD 180622} &&
\multicolumn{3}{c}{HD 201657} && \multicolumn{3}{c}{HD 201824}\\
\cline{3-5} \cline{7-9} \cline{11-13} \cline{15-17} \\
\feh && \multicolumn{3}{c}{-0.16}&&\multicolumn{3}{c}{0.21} &&
\multicolumn{3}{c}{-0.31} && \multicolumn{3}{c}{-0.40}\\
\noalign{\smallskip}
Ion &&
N & $\log\epsilon$(X) & [X/Fe]  && N &$\log\epsilon$(X) & [X/Fe] && N & $\log\epsilon$(X) & [X/Fe] &&
N & $\log\epsilon$(X) & [X/Fe]   \\
\noalign{\smallskip}
\hline
\noalign{\smallskip}
\ion{Fe}{i} &&51 &7.34 &$-$      &&35 &7.71 &$-$      &&45&7.18 &$-$      &&28&7.10   &$-$    \\
\ion{Fe}{ii}&&8  &7.32 &$-$      &&2  &7.73 &$-$      &&2 &7.31 &$-$      &&2 &7.14   &$-$    \\
\ion{Na}{i} &&2  &6.05 &$-$0.12  &&2  &6.64 &0.10     &&2 &5.99 &$-$0.03  &&2 &5.86   &$-$0.07\\
\ion{Mg}{i} &&2  &7.20 &$-$0.22  &&1  &7.78 &$-$0.01  &&1 &7.42 &0.15     &&2 &7.28   &0.10   \\
\ion{Al}{i} &&4  &6.23 &$-$0.08  &&3  &6.88 &0.20     &&3 &6.41 &0.25     &&3 &6.21   &0.14   \\
\ion{Si}{i} &&13 &7.46 &0.07     &&6  &7.97 &0.21     &&4 &7.80 &0.56     &&6 &7.36   &0.21   \\
\ion{Ca}{i} &&12 &6.23 &0.03     &&2  &6.29 &$-$0.28  &&9&5.88 &$-$0.17   &&17&6.03   &0.07   \\
\ion{Sc}{ii}&&2  &2.99 &$-$0.02  &&3  &3.37 &$-$0.01  &&3 &2.90 &0.04     &&3 &2.89   &0.12   \\
\ion{Ti}{i} &&6  &4.69 &$-$0.17  &&6  &4.98 &$-$0.25  &&6 &4.27 &$-$0.44  &&9 &4.24   &$-$0.38\\
\ion{V}{i}  &&1  &3.71 &$-$0.13  &&3  &4.05 &$-$0.16  &&1 &3.08 &$-$0.61  &&0 &$-$    &$-$    \\
\ion{Cr}{i} &&7  &5.44 &$-$0.07  &&5  &5.85 &$-$0.03  &&4 &5.21 &$-$0.15  &&4 &5.24   &$-$0.03 \\
\ion{Mn}{i} &&2  &5.10 &$-$0.13  &&2  &5.68 &0.08     &&2 &4.91 &$-$0.17  &&2 &4.98   &$-$0.01 \\
\ion{Ni}{i} &&25 &6.02 &$-$0.07  &&21 &6.63 &0.17     &&21&5.97 &0.06     &&25&5.85   &0.00   \\
\ion{Y}{i}  &&1  &2.56 &0.48     &&1  &2.90 &0.45     &&1 &2.55 &0.62     &&1 &2.30   &0.46   \\
\ion{Zr}{i} &&3  &2.88 &0.44     &&4  &3.23 &0.42     &&3 &2.84 &0.67     &&2 &2.80   &0.60   \\
\ion{Ba}{ii}&&3  &2.50 &0.53     &&3  &2.79 &0.45     &&3 &3.03 &1.21     &&3 &3.00   &1.27   \\
\ion{La}{ii}&&0  &$-$  &$-$      &&1  &2.00 &0.62     &&1 &2.29 &1.43     &&1 &2.04   &1.27   \\
\ion{Eu}{ii}&&1  &0.78 &0.43     &&1  &1.24 &0.52     &&1 &0.83 &0.63     &&1 &0.77   &0.66   \\
\noalign{\smallskip}
\hline
\noalign{\smallskip}
&& \multicolumn{3}{c}{HD 210946} && \multicolumn{3}{c}{HD 211594} &&
 \multicolumn{3}{c}{HD 216219} && \multicolumn{3}{c}{HD 223617} \\
\cline{3-5} \cline{7-9} \cline{11-13} \cline{15-17}  \\
\feh && \multicolumn{3}{c}{-0.22} && \multicolumn{3}{c}{-0.23} &&
\multicolumn{3}{c}{-0.34} && \multicolumn{3}{c}{-0.10} \\
\noalign{\smallskip}
Ion &&
N & $\log\epsilon$(X) & [X/Fe] && N & $\log\epsilon$(X) & [X/Fe] && N & $\log\epsilon$(X) & [X/Fe] &&
N & $\log\epsilon$(X) & [X/Fe]  \\
\noalign{\smallskip}
\hline
\noalign{\smallskip}
\ion{Fe}{i}  &&51&7.28 &$-$     &&49&7.27 &$-$     &&51&7.15 &$-$     &&34&7.39 &$-$ \\
\ion{Fe}{ii} &&3 &7.38 &$-$     &&0 &$-$  &$-$     &&3 &7.30 &$-$     &&4 &7.46 &$-$ \\
\ion{Na}{i}  &&1 &6.12 &0.01    &&2 &6.12 &0.02    &&2 &5.91 &$-$0.08 &&2 &6.35 &0.12 \\
\ion{Mg}{i}  &&2 &6.98 &$-$0.38 &&1 &7.15 &$-$0.2  &&3 &7.25 &0.01    &&2 &7.64 &0.16 \\
\ion{Al}{i}  &&2 &6.40 &0.15    &&2 &6.25 &0.01    &&3 &6.06 &$-$0.07 &&1 &6.35 &$-$0.02\\
\ion{Si}{i}  &&4 &7.51 &0.18    &&8 &7.65 &0.33    &&13&7.35 &0.14    &&8&7.79 &0.34 \\
\ion{Ca}{i}  &&16&5.98 &$-$0.16 &&18&6.03 &$-$0.10 &&15&6.10 &0.08    &&9&6.14 &$-$0.12 \\
\ion{Sc}{ii} &&3 &2.92 &$-$0.03 &&5 &3.04 &0.10    &&4 &2.89 &0.06    &&3 &3.11 &0.04 \\
\ion{Ti}{i}  &&6 &4.47 &$-$0.33 &&2 &4.44 &$-$0.35 &&3 &4.53 &$-$0.15 &&6 &4.67 &$-$0.25 \\
\ion{V}{i}   &&3 &3.39 &$-$0.39 &&1 &3.40 &$-$0.37 &&2 &3.61 &$-$0.05 &&2 &3.54 &$-$0.36 \\
\ion{Cr}{i}  &&5 &5.38 &$-$0.07 &&5 &5.29 &$-$0.15 &&5 &5.31 &$-$0.02 &&5 &5.71 &0.14 \\
\ion{Mn}{i}  &&2 &5.18 &0.01    &&2 &5.05 &$-$0.11 &&2 &4.88 &$-$0.17 &&2 &5.36 &0.07 \\
\ion{Ni}{i}  &&21&6.03 &0.00    &&20&6.04 &0.02    &&30&5.83 &$-$0.08 &&17&6.20 &0.05 \\
\ion{Y}{i}   &&1 &2.35 &0.33    &&1 &2.87 &0.86    &&0 &$-$  &$-$     &&1 &2.64 &0.50 \\
\ion{Zr}{i}  &&3 &3.06 &0.68    &&2 &3.08 &0.71    &&3 &3.13 &0.87    &&3 &3.15 &0.65 \\
\ion{Ba}{ii} &&3 &2.72 &0.81    &&3 &3.16 &1.26    &&3 &2.84 &1.05    &&3 &2.88 &0.85 \\
\ion{La}{ii} &&1 &1.86 &0.91    &&1 &2.18 &1.24    &&1 &1.93 &1.10    &&1 &2.11 &1.04 \\
\ion{Eu}{ii} &&1 &0.50 &0.21    &&1 &0.88 &0.60    &&1 &0.66 &0.49    &&1 &0.94 &0.53 \\
\noalign{\smallskip}
\hline\hline
\end{tabular}
   $$
}
\end{table}
\begin{table}
\tablenum{5} \caption{The detailed uncertainties of the abundance
analysis for one representative star HD 216219, and the total
uncertainties on the abundances for all other sample stars.}
{\tiny
\begin{tabular}{lrrrrrrr}
\noalign{\smallskip}
\hline
\noalign{\smallskip}
\multicolumn{8}{r}{HD 216219~$\teff=5197~\logg=3.39~\feh = $-$0.63~\xi_t=1.4$} \\
\hline
\noalign{\smallskip}
&&$\sigma_{EW}$&$\Delta \teff$ &$\Delta \logg $
& $\Delta \feh $
& $\Delta \xi_t$
& $\sigma_{tot}$\\
&      &     & +100K  & +0.2  & +0.1  & +0.2   \\
\noalign{\smallskip}
\hline
\noalign{\smallskip}
$\Delta \feh_{I}$  && 0.07 &   0.08 &$-$0.01 &   0.00 &$-$0.04 & 0.11\\[1mm]
$\Delta \feh_{II}$ && 0.06 &$-$0.03 &   0.08 &   0.03 &$-$0.04 & 0.12\\[1mm]
$\Delta \nafe$     && 0.04 &   0.06 &   0.00 &   0.00 &   0.00 & 0.07\\[1mm]
$\Delta \mgfe$     && 0.07 &   0.07 &$-$0.06 &   0.01 &$-$0.03 & 0.12\\[1mm]
$\Delta \alfe$     && 0.03 &   0.04 &$-$0.02 &$-$0.01 &$-$0.01 & 0.06\\[1mm]
$\Delta \sife$     && 0.05 &   0.03 &   0.00 &   0.00 &$-$0.02 & 0.06\\[1mm]
$\Delta \cafe$     && 0.08 &   0.08 &$-$0.04 &   0.00 &$-$0.06 & 0.13\\[1mm]
$\Delta \scfe$     && 0.09 &   0.01 &   0.06 &   0.02 &$-$0.07 & 0.13\\[1mm]
$\Delta \tife$     && 0.04 &   0.10 &$-$0.01 &$-$0.01 &$-$0.02 & 0.11\\[1mm]
$\Delta \vfe$      && 0.04 &   0.11 &$-$0.01 &$-$0.01 &$-$0.02 & 0.12\\[1mm]
$\Delta \crfe$     && 0.04 &   0.07 &$-$0.01 &$-$0.01 &$-$0.02 & 0.08\\[1mm]
$\Delta \mnfe$     && 0.07 &   0.09 &$-$0.02 &   0.00 &$-$0.05 & 0.13\\[1mm]
$\Delta \nife$     && 0.06 &   0.08 &$-$0.01 &   0.00 &$-$0.04 & 0.11\\[1mm]
$\Delta \zrfe$     && 0.04 &   0.13 &   0.00 &   0.00 &   0.00 & 0.14\\[1mm]
$\Delta \bafe$     && 0.06 &   0.04 &$-$0.02 &   0.04 &$-$0.05 & 0.10\\[1mm]
$\Delta \lafe$     && 0.04 &   0.03 &   0.08 &   0.04 &$-$0.03 & 0.11\\[1mm]
$\Delta \eufe$     && 0.03 &   0.00 &   0.08 &   0.02 &$-$0.02 & 0.09\\[1mm]
\noalign{\smallskip}
\hline
\noalign{\smallskip}
 $\sigma_{tot}$&4395 &180622&201657&201824&210946&211594&223617\\
\hline
\noalign{\smallskip}
$\Delta \feh_{I}$  &0.11  &0.14  &0.10  &0.12  &0.10  &0.11  &0.13\\[1mm]
$\Delta \feh_{II}$ &0.13  &0.24  &0.22  &0.19  &0.20  &$-$   &0.20\\[1mm]
$\Delta \nafe$     &0.08  &0.16  &0.14  &0.10  &0.13  &0.13  &0.14\\[1mm]
$\Delta \mgfe$     &0.09  &0.16  &0.14  &0.16  &0.12  &0.10  &0.13\\[1mm]
$\Delta \alfe$     &0.07  &0.13  &0.10  &0.09  &0.10  &0.09  &0.10\\[1mm]
$\Delta \sife$     &0.07  &0.14  &0.12  &0.10  &0.10  &0.11  &0.11\\[1mm]
$\Delta \cafe$     &0.15  &0.18  &0.20  &0.21  &0.18  &0.19  &0.20\\[1mm]
$\Delta \scfe$     &0.14  &0.22  &0.17  &0.19  &0.17  &0.18  &0.18\\[1mm]
$\Delta \tife$     &0.12  &0.24  &0.21  &0.20  &0.19  &0.22  &0.21\\[1mm]
$\Delta \vfe$      &0.12  &0.30  &0.23  &$-$   &0.22  &0.20  &0.25\\[1mm]
$\Delta \crfe$     &0.10  &0.21  &0.15  &0.15  &0.14  &0.14  &0.19\\[1mm]
$\Delta \mnfe$     &0.14  &0.20  &0.19  &0.24  &0.21  &0.19  &0.20\\[1mm]
$\Delta \nife$     &0.11  &0.17  &0.14  &0.19  &0.13  &0.15  &0.15\\[1mm]
$\Delta \yfe$      &0.12  &0.26  &0.24  &0.22  &0.17  &0.23  &$-$\\[1mm]
$\Delta \zrfe$     &0.13  &0.25  &0.25  &0.26  &0.19  &0.22  &0.24\\[1mm]
$\Delta \bafe$     &0.16  &0.17  &0.09  &0.09  &0.13  &0.13  &0.11\\[1mm]
$\Delta \lafe$     &$-$   &0.17  &0.20  &0.21  &0.13  &0.18  &0.17\\[1mm]
$\Delta \eufe$     &0.11  &0.15  &0.10  &0.14  &0.11  &0.12  &0.11\\[1mm]
\hline
\end{tabular}
}
\end{table}
\subsection{Microturbulence velocity}
The value of microturbulence velocity $\xi_t$ was determined from
the abundance analysis by requiring a null correlation between
[Fe/H] and the EWs. We applied this calculation with a large range
of EWs (20 $-$ 150 m{\AA}) for \ion{Fe}{i} lines. With this
selection, the uncertainty of the microturbulence velocity is
about 0.2 km\,s$^{-1}$.

Tabel~3 presents the atmospheric parameters for the sample stars,
where Column (1)-(6) list, consequently, the HD identifications,
\teff, log$g$, mass, microturbulence velocity and [Fe/H]. The
temperature coverage of the stars are from 4284 to 5553 K, the
surface gravity coverage are from 1.67 to 3.64. The
microturbulence velocity is from 1.3 to 1.7, and their [Fe/H] is
from $-$0.40 to 0.21. The uncertainties of the parameters are:
$\sigma(\teff) = 100$~K, $\sigma(\logg) = 0.2$, $\sigma(\feh) =
0.1$, and $\sigma(\xi_{t})=0.2$ km\,s$^{-1}$.

\begin{figure}
\centering \hspace{3mm} \psfig{figure=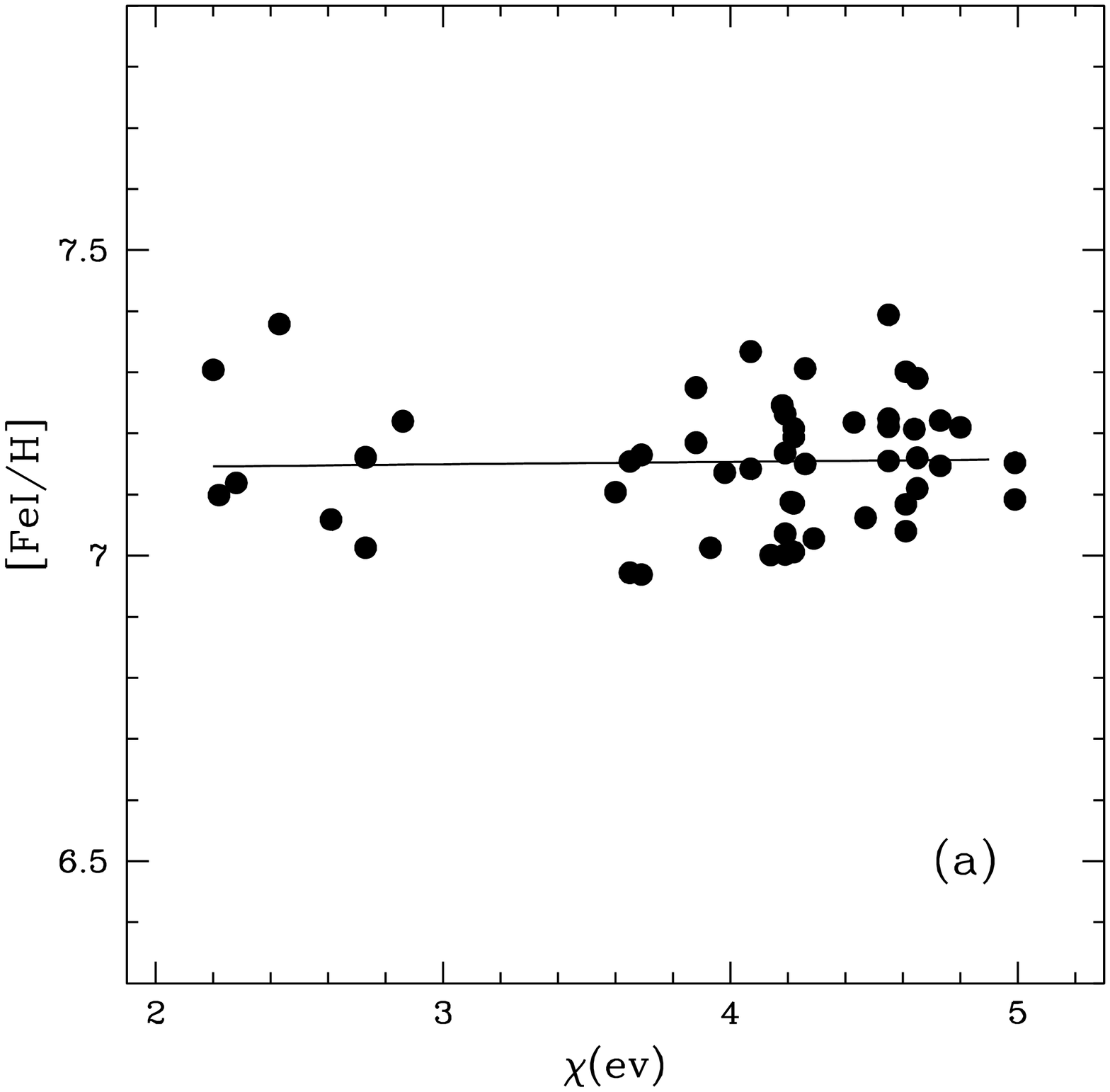,width=50mm}
\psfig{figure=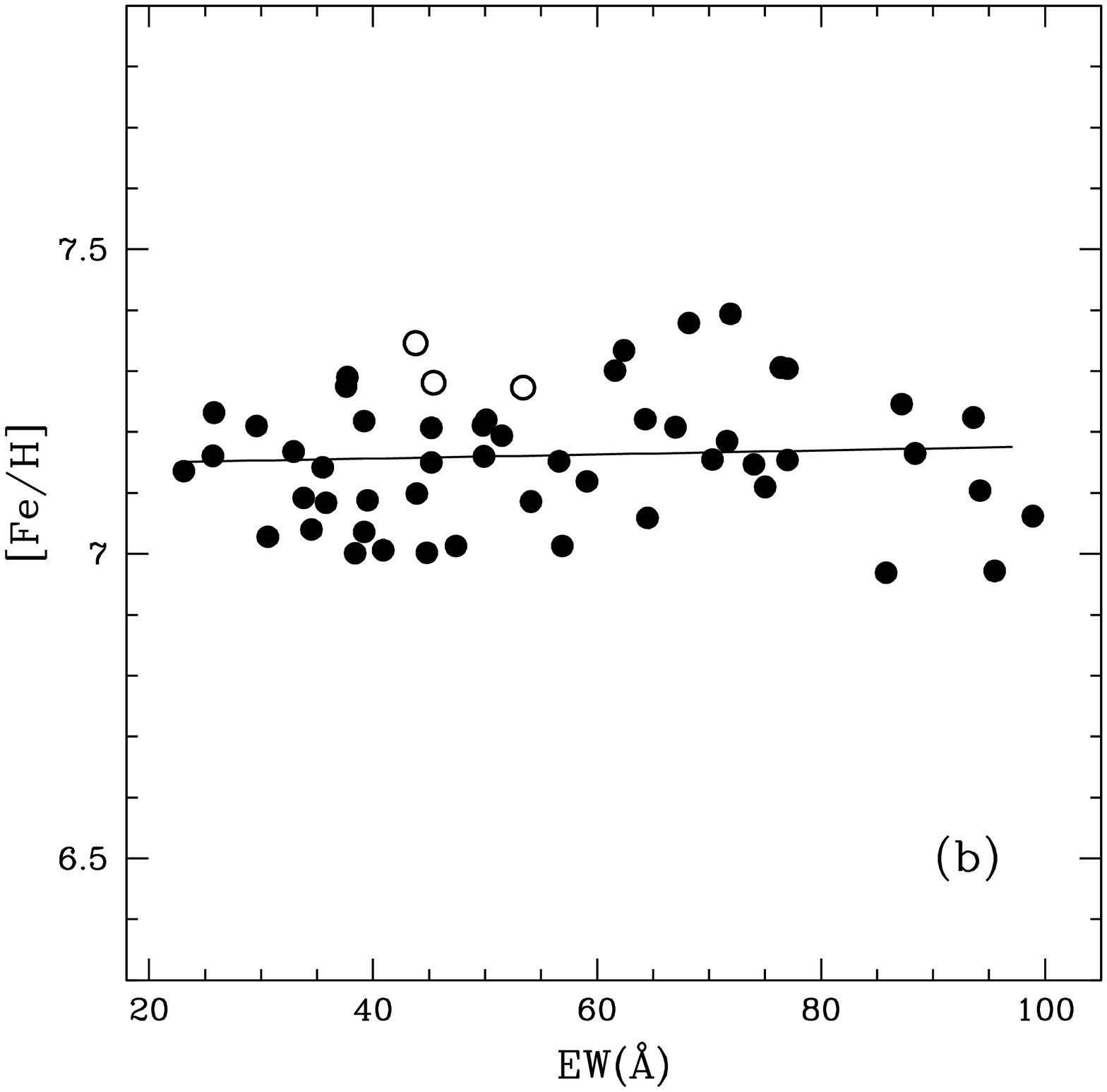,width=50mm} \caption{To check the
reliability of the determined atmospheric parameters
$\teff$/log$g$/$\xi_t$ of the stars by taking HD216219 as an
example: (a). Fe abundances from Fe\,I lines as a function of
excitation potential. There is no significant trend of [Fe/H] with
$\chi$, indicating a correct temperature distribution in the model
atmosphere. (b). The consistence of the Fe\,I and Fe,II abundances
(the filled and open circles refer to the Fe\,I and Fe\,II
abundances respectively), which mean that the determined log$g$
are reliable; and an illustration of the determination of the
microturbulence velocity since there is no significant trend
between [Fe/H] and EWs.} \label{figCheck}
\end{figure}

The reliability of the derived atmospheric parameters
$\teff$/log$g$/$\xi_t$ are confirmed by the further checks. Taking
HD\,216219 as a representative,  Figure~\ref{figCheck}a give the Fe
abundances from different Fe\,I lines as a function of their
excitation potential, which fulfills the excitation equilibrium;
Figure~\ref{figCheck}b shows that the Fe abundances from Fe\,I lines
and Fe\,II lines are consistent within 0.2\,dex, which illustrates
the ionization equilibrium, and also shows that there is no trend
between Fe abundances and EWs of the lines.

Comparing our results with those of Smith et al. (1993), the
derived atmospheric parameters for HD\,4395 are $T_{\rm
eff}$=5447/5450K, log$g$=3.60/3.3, $\xi_t$=1.3/1.3,
[Fe/H]=-0.16/-0.33; for HD\,216219 are $T_{\rm eff}$=5553/5600K,
log$g$=3.64/3.2, $\xi_t$=1.4/1.6, [Fe/H]=-0.34/-0.32. They are
well consistent.

\section{Stellar atmosphere model and spectral lines}
The stellar atmospheric model is implemented by ATLAS9 code
(Kurucz \cite{kuru93}) to do the abundance analysis. This is LTE,
plane-parallel, line-blanketed models. Abundances of chemical
elements were determined by using the input atmospheric parameters
given in Table~3 and the measured EWs of the absorption lines. All
the lines adopted in determining element abundances are presented
in Table~2, which shows the spectral lines and wavelengths,
excitation potential $\chi$, oscillator strengths log\,$gf$, EWs
and log$\epsilon$ of each line. The selections of the
lines have been described in Sect.2. The oscillator strengths
log$gf$ of spectral lines are taken from the NIST database
(http://physics.nist.gov), Lambert \& Warner (1968), Weise \&
Martin (1980), Bi$\acute{e}$mont et al. (1981, 1982), Hannaford et
al. (1982), Fuhr et al. (1988), Luck \& Bond (1991), O'Brian et
al. (1991), Bard \& Kock (1994), Lambert et al. (1996), Nissen \&
Schuster (1997), Chen et al. (2000), Liang et al. (2003) and the
references therein. Col.\,(5) of Table\,2 gives these reference
sources for the spectral lines.
\section{Chemical abundances and analysis}
In this section, we present the determined abundances of the
sample stars for about 20 elements based on the spectral
observations and atmospheric model.

\subsection{Abundance of barium stars}
The derived element abundances of all the sample stars are given
in Table~4, including log\,$\epsilon$ and the corresponding
[$X$/Fe] values for all ions. The solar abundances are adopted
from Grevesse \& Sauval (1998).

Figure~\ref{abun} directly presents the abundance results of our
sample stars, including the chemical elements, Na, Mg, Al, Si, Ca,
Sc, Ti, V, Cr, Mn, Ni, Y, Zr, Ba, La and Eu. It is obvious to show
that the neutron capture process elements, Y, Zr, Ba, La, Eu, are
overabundant than the solar abundances. Especially, Y and Zr
exhibit as the first peak while Ba and La exhibit as the second
peak, and the second peak is higher than the first one. Other
elements from Na to Ni, such as $\alpha$ elements and iron
elements, show similar abundances to the solar, which means that
these Ba stars belong to disk stars. The behaviors of Sc and Mn
are compatible to the results of Nissen et al. (2000) and Chen et
al. (\cite{cnzszb00}), who demonstrated that decreasing [Sc/Fe]
with increasing mentallicity in disk stars, whereas [Mn/Fe]
increases with increasing [Fe/H].

\subsection{Uncertainties in the abundances}
There are two kinds of uncertainties in the abundance
determination: the systematic errors introduced by the atmospheric
parameters, and the random errors in determining EWs, oscillator
strengths, and damping constants. We ignore the uncertainties in
atomic data since they could be small (Chen et al. \cite{cnzzb00})
and consider the uncertainties in atmospheric parameter
determinations and EW measurements. Assuming that the effects of
the uncertainties of the parameters are independent, we can
estimate the total uncertainty with Eq.~(\ref{delta}):
\begin{equation}
\label{delta} \sigma_{total} = \sqrt{(\sigma_{EW})^2
+ (\Delta \teff)^2 + (\Delta \logg)^2 + (\Delta \feh)^2 + (\Delta
\xi_t)^2},
\end{equation}
where $\sigma_{EW}$, $\Delta \teff$, $\Delta \logg$, $\Delta
\feh$ and $\Delta \xi_t$ are the corresponding variations in the
ion abundances due to the variations on equivalent widths, \teff,
\logg, mental abundance and microturbulent velocity,
respectively.

For our spectra, the typical uncertainty of the EW is about
6.1\%. ~Table~5 shows the effects on the derived abundances
changed by 6.1\% in EW, 100 K in effective temperature, 0.2 dex
in surface gravity, 0.1 dex in metallicity, and 0.2 km\,s$^{-1}$
in microturbulence velocity for one representative star,
HD\,216219. The total uncertainties on the output abundances have
also be given in Table~5 for all other sample stars by considering
the same errors as above in the individual atmospheric parameter.

\subsection{Comparisons with Smith et al. (1993)}
\begin{figure}
\centering \hspace{3mm} \psfig{figure=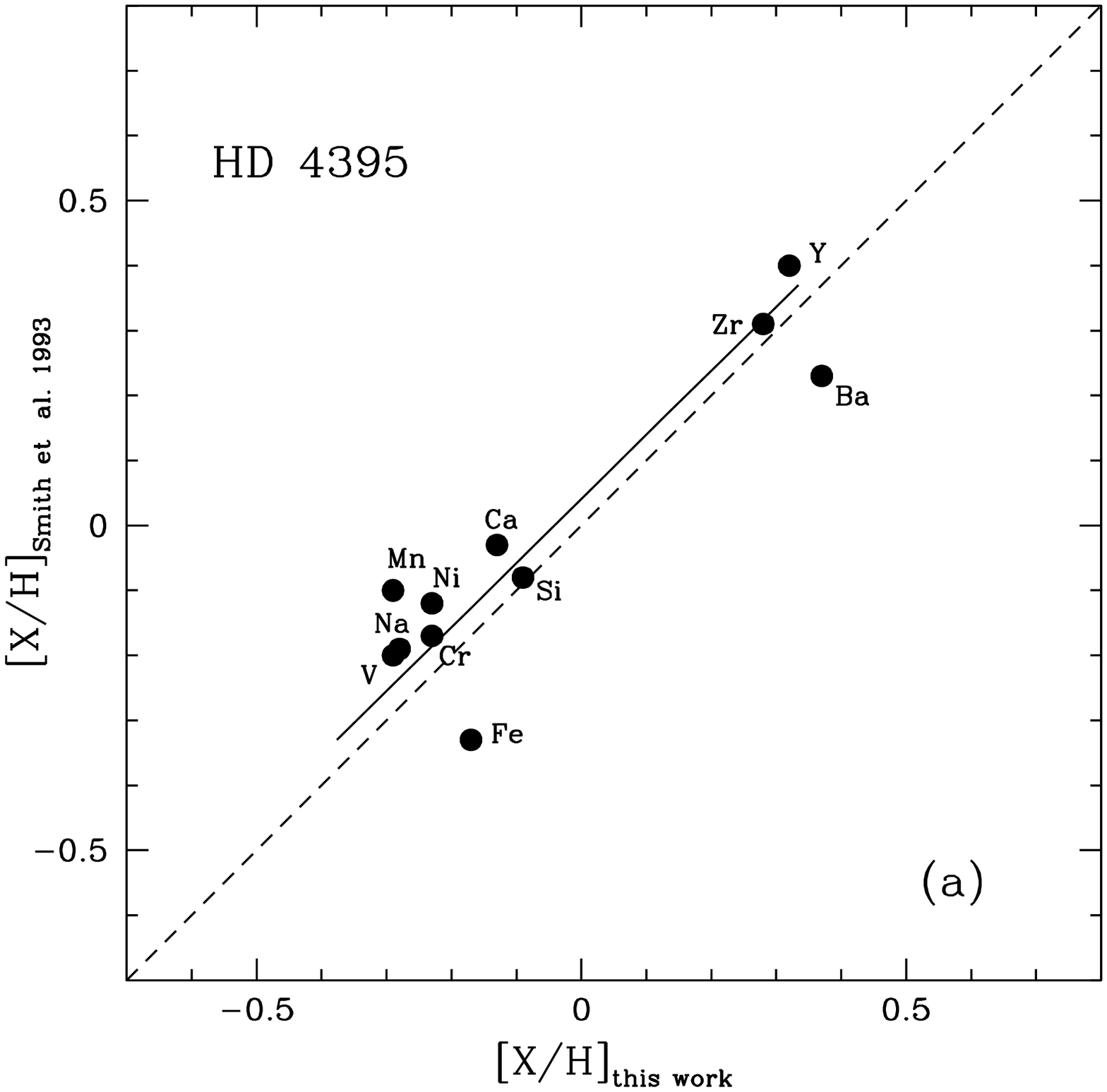,width=50mm}
\psfig{figure=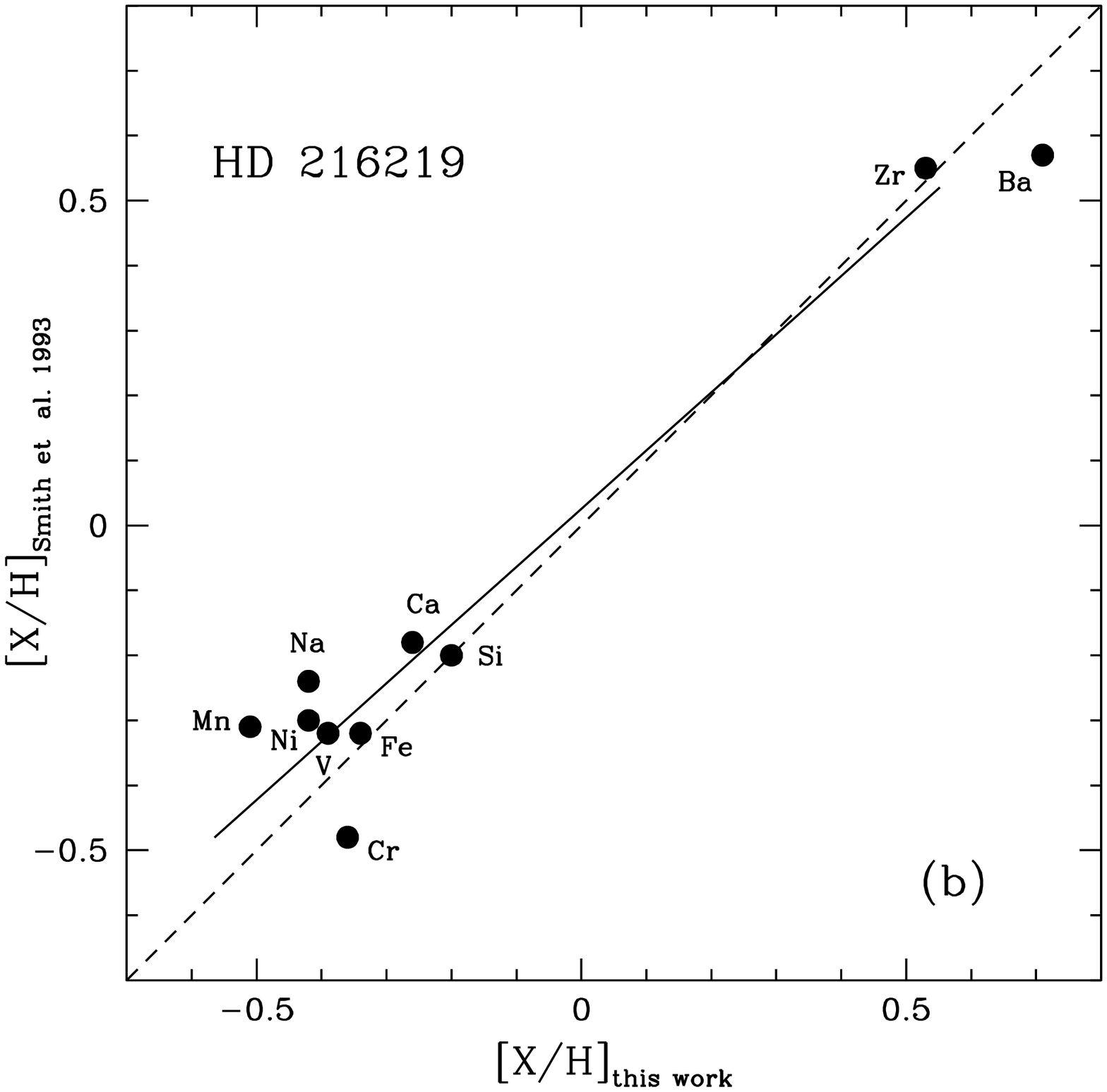,width=50mm} \caption{The
comparisons between our abundances determinations and those of
Smith et al. (1993) for the two common stars: (a). for HD\,4395;
(b). for HD\,216219. The solid lines are the least square fits for
the data and the dashed lines are the one-to-one relations. Our
results are consistent with theirs.}
\label{comAbun}
\end{figure}

As for the two common stars with Smith et al. (1993), we compare
our abundance determinations with theirs for HD\,4395 and
HD\,216219. Figure~\ref{comAbun} shows the consistences between our
abundance estimations with theirs are within 0.2\,dex, but most
are in 0.1\,dex.

%
%
%
\section{Comparing with wind accretion model results}

We try to use the wind accretion model to predict the theoretical
heavy element abundances of Ba stars in binary systems, and then
compare these theoretical predicts with the observed abundance
patterns of our sample stars. Following Liang et al. (2000, 2003),
the calculations of theoretical abundances are made in two steps:
the AGB nucleosynthesis based on the latest TP-AGB model and the
branch path of the $s$-process nucleosynthesis, and the binary
accretion based on the angular momentum conservation model of wind
accretion.

The standard case of our wind accretion model is:
$M_{1,0}$=3.0$M_{\odot}$, $M_{2,0}$=1.3$M_{\odot}$, $v_{\rm
ej}$=15${\rm ~km~s^{-1}}$ ($M_{1,0}$ is the main sequence mass of
the intrinsic AGB star, the present white dwarf, in the binary
system; $M_{2,0}$ is the corresponding mass of the present Ba
star; $v_{\rm ej}$ is the wind velocity). We assumed the standard
accretion rate is 0.15 times of the Bondi-Hoyle's accretion rate
(Liang et al. \cite{lzz00}; Boffin \& Za$\check{c}$s~\cite{bz94}).
The observed orbital elements of the sample stars are listed in
Table~6, which are taken from Jorissen et al. (\cite{jvmu98}) and
Preston \& Sneden (\cite{ps01}). The observed orbital periods of
our sample stars cover from 1018.9 to 6200 days and the
eccentricities range from 0.06 to 0.65.
\begin{table}
\tablenum{6}
\caption{Orbital elements of the sample stars derived from literatures.}
\begin{tabular}{llllrr}
\hline\hline
\noalign{\smallskip}
HD& $P$ & $e$ & Classes & Reference \\
  &(days) &       \\
\hline
\noalign{\smallskip}
4395  & 6200  &0.65& $-$  &Preston \& Sneden \cite{ps01}\\
180622& 4049.2&0.06&mild  &Jorissen et al. \cite{jvmu98}\\
201657& 1710.4&0.17&strong&Jorissen et al. \cite{jvmu98}\\
201824& 2837  &0.34&strong&Jorissen et al. \cite{jvmu98}\\
210946& 1529.5&0.13&mild  &Jorissen et al. \cite{jvmu98}\\
211594& 1018.9&0.06&strong&Jorissen et al. \cite{jvmu98}\\
216219& 4098.0&0.10&mild  &Jorissen et al. \cite{jvmu98}\\
223617& 1293.7&0.06&mild  &Jorissen et al. \cite{jvmu98}\\
\noalign{\smallskip}
\hline
\end{tabular}
\end{table}

\begin{figure}
\centering
\includegraphics{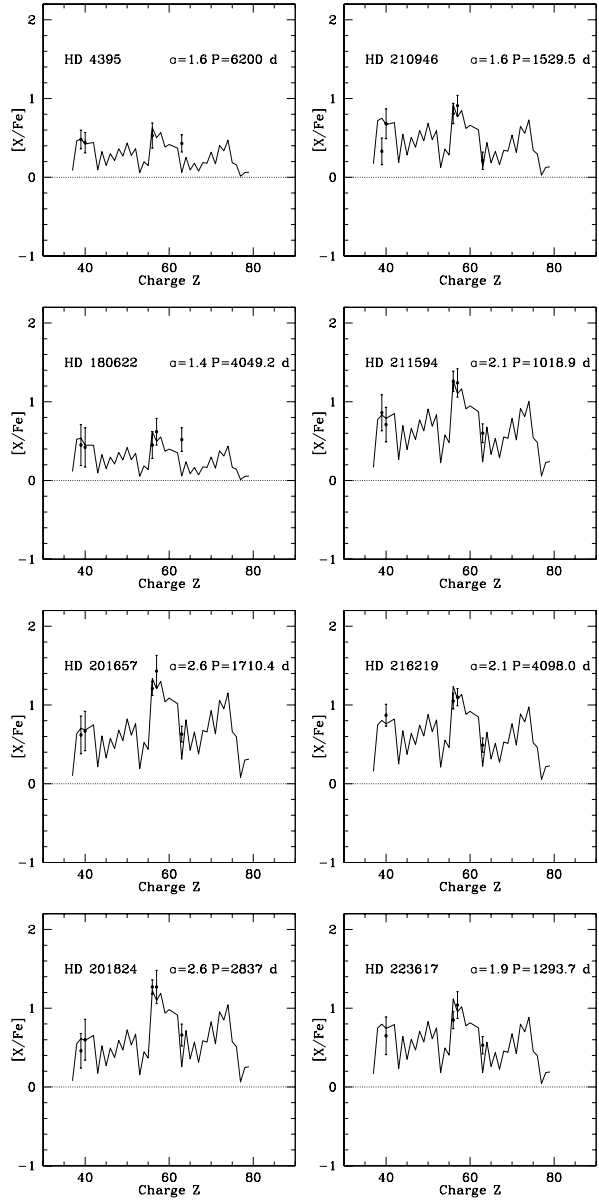}
\caption{The fitting of the theoretical to observed heavy-element
abundances of sample stars with standard case of wind
accretion. Label ``$a$" represents the times of the corresponding
standard neutron exposures in the $^{13}$C profile in the AGB
progenitor suggested by Gallino et al. (1998). $P$ refers to the
orbital period of the sample star.} \label{fitting}
\end{figure}

Figure~\ref{fitting} shows the comparisons between the theoretical
abundances from the wind accretion model and the observed
abundances of the sample stars, and the uncertainties of the
observed abundances are marked as well. In the figure, the
variable ``$a$" represents the times of the corresponding standard
neutron exposure in the $^{13}$C profile in the AGB progenitor
companion suggested by Gallino et al. (1998) and the higher $a$
value reflects the higher neutron exposure occurred in interiors
of the AGB progenitor. $P$ refers to the orbital period of the
sample star.

There is good agreement between our observed abundances and the
theoretical ones for the sample stars. These mean that wind
accretion can be the formation scenario of these Ba stars in
binary systems. These are consistent with the suggestions of
Jorissen et al. (1998), Zhang et al. (1999) and Liang et al.
(2000), who mentioned that Ba stars with periods longer than 1500
or 1600 days could be formed through wind accretion. We
should notice that the heavy element abundance patterns of two
sample stars with $P>1000$ days, HD\,211594 and HD\,223617, can
also been explained by wind accretion. Figure~\ref{fitting} also
shows that the strong Ba stars generally require the higher ``$a$"
values in the model than the mild Ba stars, i.e., the stronger
neutron exposure occurred in the AGB progenitors in their
$s$-process nucleosynthesis.

\section{Discussions and conclusions}
\label{sec.7}

The chemical compositions of six Ba stars and two CH subgiant
stars were obtained on the basis of the high S/N ratio and high
resolution spectra observed by using the 2.16m telescope at
NAOC/Xinglong station. Their stellar atmospheric parameters were
determined from the reliable methods, and show the ranges of 4284
$<$ \teff $<$ 5553, 1.67 $<$ log$g$ $<$ 3.64, $-$0.40 $<$ \feh $<$
0.21, and 1.3 $<$ $\xi_t$ $<$ 1.7. The model atmospheres were
generated by using ATLAS9 code and the updated atomic data of the
selected spectral lines for measuring EWs.

Then we obtain the abundances of chemical elements, Na, Mg, Al,
Si, Ca, Sc, Ti, V, Cr, Mn, Ni, Y, Zr, Ba, La, Eu for our eight
sample stars. The elements from Na to Ni, such as $\alpha$ and
iron peak elements, show comparable abundances to the Sun,
associated with the \feh~in a range of $-$0.40 to 0.21,
which mean these Ba stars belong to the Galactic disk. The neutron
capture process elements Y, Zr, Ba, La, Eu show obvious
overabundances than the solar abundances, for example, their
[Ba/Fe] values are from 0.45 to 1.27. The abundance patterns of
our sample stars are consistent with those obtained for other Ba
stars in Za$\check{c}$s (\cite{z94}), Liang et al. (2003),
Smiljanic et al. (\cite{sps07}). Our study enlarge the sample
of Ba stars with known chemical abundances. And we further check
the formation scenario of these sample stars through theoretical
wind accretion model.

We adopt the angular momentum conservation model of wind accretion
to calculate the chemical abundances of Ba stars in the binary
systems. The predicted results by the model can explain well the
observed abundance patterns of the $s$-process elements. The
abundance patterns of two sample stars, HD\,211594 and HD\,223617,
can also been explained by the wind accretion model although their
orbital periods are 1018.9 and 1293.7 days respectively, which are
lower than the low limit of wind accretion formation of Ba stars
suggested by Jorissen et al. (\cite{jvmu98}), 1500 days, and Zhang
et al. (\cite{zllp99}), Liang et al. (\cite{lzz00}), 1600 days.
This result could further decrease such low limit to be about 1000
days.

The masses of the sample stars are also determined and given in
Table~3, as well as their errors. For most of them, their masses
(0.78-1.92$M_{\odot}$) are close to the average masses of typical
mild and strong Ba stars given in Jorissen et al. (\cite{jvmu98})
(their Table\,9), who suggested that the average mass of typical
mild Ba stars is 1.9 or 2.3$M_{\odot}$ with the 0.60 or
0.67$M_{\odot}$ companion white dwarfs, and the average mass of
typical strong Ba stars is 1.5 or 1.9$M_{\odot}$ with the 0.60 or
0.67$M_{\odot}$ companion white dwarfs. However, the very high
mass (4.58$M_{\odot}$) of HD\,201824 should not be real and the
reason could be the big error of its parallax, up to 3 times
(1.56/0.56) uncertainty, which will cause the uncertainties of
$\sim$0.4\,dex in log$g$ and 3.2$M_{\odot}$ in mass. This 0.4\,dex
uncertainty in log$g$ is larger than the general case (0.2\,dex)
of the sample stars, but will not affect much the abundances. In
Table\,5 and Figure~\ref{fitting}, we adopt the general
uncertainties of log$g$, 0.2\,dex, to estimate the uncertainties
on abundances for HD\,201824.

 The derived abundances of our sample stars confirm well their
``strong" or ``mild" Ba star properties.As shown in Table 4, Figure~\ref{abun}
and Figure~\ref{fitting}, for the four mild Ba stars, namely, HD\,180622,
HD\,210946, HD\,216219 and HD\,223617, their average abundances of
$s$-process elements are [Ba/Fe]=0.79, [La/Fe]=0.92, [Y/Fe]=0.43,
[Zr/Fe]=0.66, [Eu/Fe]=0.44; and for the three strong Ba stars,
HD\,201657, HD\,201824 and HD\,211594, their average abundances of
$s$-process elements are [Ba/Fe]=1.25, [La/Fe]=1.31, [Y/Fe]=0.65,
[Zr/Fe]=0.66, [Eu/Fe]=0.63. These show that the s-process
element abundances of strong Ba stars are about 0.4\,dex (or
0.2\,dex) higher than those of the mild Ba stars. However, there
are no obvious trend to show that the strong and mild Ba stars
have different ranges in metallicity (also see Smiljanic et al.
\cite{sps07}). Also, there are no obvious indication to show that
the orbital periods of mild Ba stars are longer than those of the
strong Ba stars.

HD\,4395 is a CH subgiant star. HD\,216219 has also been
classified as a CH subgiant (Smith et al. 1993 and the reference
therein) or a mild Ba stars (Jorissen et al. 1998;
L\"{u} et al. 1991). CH subgiant was firstly discovered by Bond
(1974). As Luck \& Bond (1982) discussed, some of their CH
subgiants might more properly be called subgiant barium stars or
even main-sequence barium stars. We did not find obvious
difference in the abundance patterns of our CH subgiant sample
stars and the Ba sample stars except HD\,4395 shows a bit
relatively lower overabundances in its s-process elements.
Moreover, the wind accretion models for binary system can explain
well the observed overabundances of s-process elements in the CH
subgiant stars as for other Ba stars. However, our results are not
enough to check the suggested evolutionary relation between CH
subgiants and classical Ba stars, which will need C/O and Li
abundances (Smith et al. 1993; Lambert et al. 1993).

%


%

\begin{acknowledgements}
We thank our referee for very valuable comments which help us
a lot to improve this work.
 We would like to thank Prof. Gang Zhao, Yuqin Chen, Jianrong
Shi, Yujuan Liu, Shu Liu, Kefeng Tan and the {\sl Stellar Abundance
and Galactic Evolution Group} at NAOC for sharing their programs for
abundance analysis and the helpful discussions about data reduction
and abundance analysis. This work was supported by the Natural
Science Foundation of China (NSFC) Foundation under No.10403006,
10433010, 10673002, 10573022, 10333060, and 10521001; and the
National Basic Research Program of China (973 Program)
No.2007CB815404, 2007CB815406, 2009CB824800.

\end{acknowledgements}
%

%
%
%
%
%
\clearpage

\begin{deluxetable}{lllrrrrrrrrrrrrrrrrrr}
\tabletypesize{\scriptsize}
\rotate
\tablenum{2}

\tablecaption{Equivalent widths and abundances for all the sample stars.}
\tablewidth{0pt}
\tablehead{
\colhead{$\lambda$ (\AA)}  &  \colhead{Ion} &  \colhead{$\chi$} &  \colhead{log $gf$} & \colhead{Ref.}
& \multicolumn{2}{c}{HD 4395}&\multicolumn{2}{c}{HD 180622} &\multicolumn{2}{c}{HD 201657}&\multicolumn{2}{c}{HD 201824}&
\multicolumn{2}{c}{HD 210946}&\multicolumn{2}{c}{HD 211594}&\multicolumn{2}{c}{HD 216219}&\multicolumn{2}{c}{HD 223617}
\\
\colhead{}& &&&
& \colhead{EW} & \colhead{$\log\epsilon$} &\colhead{EW} & \colhead{$\log\epsilon$}
& \colhead{EW}& \colhead{$\log\epsilon$} & \colhead{EW}  &
 \colhead{$\log\epsilon$}&  \colhead{EW}& \colhead{$\log\epsilon$}&  \colhead{EW}& \colhead{$\log\epsilon$}
  & \colhead{EW}& \colhead{$\log\epsilon$}  & \colhead{EW} & \colhead{$\log\epsilon$}

 }
\startdata
5522.454 &Fe I & 4.21 &$-$1.55 &nd& 38.1 &7.281& 99.3 &7.916&  73.9 &7.231& $-$ & $-$ & 76.1 &7.385& 88.6 &7.600&  $-$ &  $-$& 84.4&7.575\\
5525.552 &Fe I & 4.23 &$-$1.08 &cn&  $-$ &$-$  &  $-$ &$-$  &  87.0 &7.025& $-$ & $-$ &  $-$ & $-$ &  $-$ &  $-$&  $-$ &  $-$&  $-$&  $-$\\
5543.944 &Fe I & 4.22 &$-$1.14 &nd& 69.5 &7.490&  $-$ &$-$  &   $-$ & $-$ & $-$ & $-$ &  $-$ & $-$ & 93.5 &7.293& 51.5 &7.194&  $-$&  $-$\\
5546.514 &Fe I & 4.37 &$-$1.31 &nd& 49.9 &7.431&  $-$ &$-$  &   $-$ & $-$ & $-$ & $-$ & 74.6 &7.310& 82.0 &7.430&  $-$ &  $-$&  $-$&  $-$\\
5560.220 &Fe I & 4.43 &$-$1.19 &nd& $-$  &$-$  & 86.2 &7.561&  62.0 &6.942&  $-$&  $-$& 65.3 &7.093& 57.9 &6.947& 39.2 &7.218& 72.5&7.249\\
5618.642 &Fe I & 4.21 &$-$1.28 &cn& 57.2 &7.371& 99.6 &7.638&  73.2 &6.940&  $-$&  $-$& 77.2 &7.123&  $-$ & $-$ & 39.5 &7.088&  $-$&  $-$\\
5633.953 &Fe I & 4.99 &$-$0.27 &nd&  $-$ &$-$  &  $-$ & $-$ &  95.5 &7.302&  $-$&  $-$& 79.8 &7.091&  $-$ & $-$ & 56.6 &7.152&  $-$&  $-$\\
5638.271 &Fe I & 4.22 &$-$0.87 &nd&  $-$ &$-$  &  $-$ & $-$ &   $-$ &$-$  &  $-$&  $-$& 97.6 &7.113&  $-$ & $-$ & 67.0 &7.208& 97.3&7.154\\
5641.448 &Fe I & 4.26 &$-$1.18 &nd&  $-$ &$-$  &  $-$ & $-$ &  99.3 &7.379& 89.2&7.256&  $-$ &  $-$& 91.5 &7.336& 45.2 &7.150&  $-$&  $-$\\
5646.686 &Fe I & 4.26 &$-$2.51 &nd&  $-$ &$-$  & 37.9 &7.740&  18.8 &7.191& 16.0&7.093& 16.1 &7.205&  $-$ &  $-$&  $-$ &  $-$&  $-$&  $-$\\
5651.470 &Fe I & 4.47 &$-$2.00 &nd&  $-$ &$-$  & 52.0 &7.763&   $-$ & $-$ &  $-$& $-$ &  $-$ & $-$ & 44.7 &7.572&  $-$ &  $-$&  $-$&  $-$\\
5655.183 &Fe I & 5.06 &$-$0.64 &nd&  $-$ &$-$  &  $-$ & $-$ &   $-$ & $-$ &  $-$&  $-$&  $-$ & $-$ & 79.8 &7.533&  $-$ &  $-$&  $-$&  $-$\\
5662.524 &Fe I & 4.18 &$-$0.57 &cn&  $-$ &$-$  &  $-$ & $-$ &   $-$ & $-$ &  $-$&  $-$&  $-$ & $-$ &  $-$ &  $-$& 87.2 &7.246&  $-$&  $-$\\
5679.032 &Fe I & 4.65 &$-$0.92 &nd&  $-$ &$-$  &  $-$ & $-$ &  71.5 &7.113&  $-$&  $-$& 76.4 &7.280& 71.5 &7.182&  $-$ &  $-$& 82.8&7.436\\
5705.473 &Fe I & 4.30 &$-$1.36 &cn& 38.8 &7.181& 84.7 &7.524&  60.1 &6.904&  $-$&  $-$& 66.2 &7.109&  $-$ &  $-$&  $-$ &  $-$& 70.2&7.200\\
5717.841 &Fe I & 4.28 &$-$1.13 &nd&  $-$ &$-$  &  $-$ & $-$ &   $-$ &$-$  & 93.1&7.302&  $-$ & $-$ &  $-$ &  $-$&  $-$ &  $-$&  $-$&  $-$\\
5731.772 &Fe I & 4.26 &$-$1.30 &nd&  $-$ &$-$  &  $-$ & $-$ &   $-$ &$-$  & 83.5&7.250&  $-$ & $-$ & 98.4 &7.579&  $-$ &  $-$& 94.3&7.564\\
5753.132 &Fe I & 4.26 &$-$0.76 &nd&  $-$ &$-$  &  $-$ & $-$ &   $-$ &$-$  &  $-$& $-$ &  $-$ &  $-$&  $-$ &  $-$& 76.4 &7.306&  $-$&  $-$\\
5775.088 &Fe I & 4.22 &$-$1.17 &cn& 63.3 &7.380&  $-$ & $-$ &  78.6 &6.931& 77.6&6.945& 78.4 &7.037& 70.9 &6.885& 40.9 &7.006& 85.9&7.210\\
5806.732 &Fe I & 4.61 &$-$1.05 &lz&  $-$ &$-$  & 94.3 &7.786&  68.8 &7.143&  $-$& $-$ & 76.1 &7.350& 66.4 &7.167&  $-$ &  $-$& 79.7&7.451\\
5809.224 &Fe I & 3.88 &$-$1.84 &lz&  $-$ &$-$  &  $-$ &$-$  &  93.6 &7.432& 68.6&7.020& 87.0 &7.451& 84.9 &7.390& 37.6 &7.275& 96.2&7.663\\
5811.917 &Fe I & 4.14 &$-$2.43 &nd&  $-$ &$-$  & 45.2 &7.647&  22.0 &7.045&  $-$&  $-$& 29.7 &7.322&  $-$ &  $-$&  $-$ &  $-$&  $-$&  $-$\\
5845.270 &Fe I & 5.03 &$-$1.82 &nd&  $-$ &$-$  & 24.5 &7.689&   $-$ & $-$ &  $-$&  $-$& 13.2 &7.316&  $-$ &  $-$&  $-$ &  $-$&  $-$&  $-$\\
5849.682 &Fe I & 3.69 &$-$2.99 &nd&  $-$ &$-$  & 54.9 &7.825&   $-$ & $-$ &  $-$&  $-$&  $-$ &  $-$&  $-$ &  $-$&  $-$ &  $-$&  $-$&  $-$\\
5852.228 &Fe I & 4.55 &$-$1.33 &nd&  $-$ &$-$  &  $-$ &$-$  &  59.2 &7.180& 42.7&6.866& 67.4 &7.400&  $-$ &  $-$&  $-$ &  $-$& 82.7&7.712\\
5856.096 &Fe I & 4.29 &$-$1.33 &ow&  $-$ &$-$  & 83.6 &7.452&  70.4 &7.037&  $-$&  $-$&  $-$ & $-$ &  $-$ &  $-$& 30.6 &7.028& 72.5&7.195\\
5858.779 &Fe I & 4.30 &$-$2.26 &nd&  $-$ &$-$  &  $-$ & $-$ &  40.7 &7.467&  $-$&  $-$&  $-$ & $-$ &  $-$ &  $-$&  $-$ &  $-$&  $-$&  $-$\\
5859.596 &Fe I & 4.55 &$-$0.66 &lz&  $-$ &$-$  &  $-$ & $-$ &   $-$ & $-$ &  $-$&  $-$&  $-$ & $-$ &  $-$ &  $-$& 71.9 &7.394&  $-$&  $-$\\
5862.368 &Fe I & 4.55 &$-$0.45 &lz&  $-$ &$-$  &  $-$ & $-$ &  94.1 &6.912&  $-$&  $-$&  $-$ & $-$ & 87.6 &6.871& 70.3 &7.155&  $-$&  $-$\\
5905.680 &Fe I & 4.65 &$-$0.73 &lz&  $-$ &$-$  &  $-$ & $-$ &  71.2 &6.914&  $-$&  $-$&  $-$ & $-$ & 66.7 &6.897& 49.9 &7.160& 75.6&7.096\\
5927.797 &Fe I & 4.65 &$-$1.09 &lz&  $-$ &$-$  & 81.3 &7.617&  50.0 &6.906&  $-$&  $-$&  $-$ & $-$ & 45.6 &6.884& 37.7 &7.290&  $-$&  $-$\\
5929.682 &Fe I & 4.55 &$-$1.41 &lz&  $-$ &$-$  &  $-$ &$-$  &  59.8 &7.269& 57.3&7.206& 56.4 &7.283& 60.7 &7.349&  $-$ &  $-$& 63.5&7.426\\
5930.191 &Fe I & 4.65 &$-$0.23 &lz&  $-$ &$-$  &  $-$ & $-$ &   $-$ & $-$ &  $-$&  $-$&  $-$ &  $-$&  $-$ &  $-$& 75.0 &7.110&  $-$&  $-$\\
5934.665 &Fe I & 3.93 &$-$1.17 &lz& 77.2 &7.352&  $-$ & $-$ &   $-$ & $-$ &  $-$&  $-$&  $-$ &  $-$&  $-$ &  $-$& 56.9 &7.013&  $-$&  $-$\\
5952.726 &Fe I & 3.98 &$-$1.44 &nd&  $-$ &$-$  &  $-$ & $-$ &   $-$ & $-$ &  $-$&  $-$&  $-$ &  $-$& 81.0 &7.033&  $-$ &  $-$&  $-$&  $-$\\
6003.022 &Fe I & 3.88 &$-$1.12 &lz& 90.4 &7.497&  $-$ & $-$ &   $-$ & $-$ &  $-$&  $-$&  $-$ &  $-$&  $-$ &  $-$& 71.6 &7.185&  $-$&  $-$\\
6024.068 &Fe I & 4.55 &$-$0.12 &nd&  $-$ &$-$  &  $-$ & $-$ &   $-$ & $-$ &  $-$&  $-$&  $-$ &  $-$&  $-$ &  $-$& 93.6 &7.224&  $-$&  $-$\\
6027.059 &Fe I & 4.07 &$-$1.09 &cn& 67.0 &7.208&  $-$ & $-$ &   $-$ & $-$ &  $-$&  $-$& 94.8 &7.062& 91.9 &6.991&  $-$ &  $-$& 98.0&7.165\\
6034.033 &Fe I & 4.31 &$-$2.42 &nd&  $-$ &$-$  & 45.2 &7.847&   $-$ & $-$ &  $-$&  $-$&  $-$ &  $-$&  $-$ &  $-$&  $-$ &  $-$&  $-$&  $-$\\
6056.013 &Fe I & 4.73 &$-$0.46 &lz& 71.6 &7.316&  $-$ & $-$ &  88.8 &7.044& 89.7&7.088& 90.0 &7.141& 80.8 &6.969& 64.3 &7.221& 94.5&7.275\\
6079.016 &Fe I & 4.65 &$-$1.12 &lz&  $-$ &$-$  & 86.8 &7.746&  68.6 &7.251&  $-$& $-$ & 73.3 &7.404& 66.5 &7.278&  $-$ &  $-$& 77.2&7.507\\
6093.649 &Fe I & 4.61 &$-$1.50 &lz& 26.2 &7.355& 71.6 &7.787&  49.4 &7.248& 38.6&7.022& 50.9 &7.344& 45.9 &7.247&  $-$ &  $-$& 58.9&7.500\\
6094.370 &Fe I & 4.65 &$-$1.94 &nd&  $-$ &$-$  &  $-$ & $-$ &   $-$ & $-$ &  $-$&  $-$&  $-$ & $-$ & 32.1 &7.472&  $-$ &  $-$&  $-$&  $-$\\
6096.671 &Fe I & 3.98 &$-$1.93 &lz& 41.9 &7.471& 83.1 &7.638&  62.4 &7.091&  $-$&  $-$& 65.7 &7.263& 54.4 &7.051& 23.1 &7.136& 69.9&7.350\\
6105.150 &Fe I & 4.55 &$-$2.07 &nd&  $-$ &$-$  &  $-$ &$-$  &   $-$ & $-$ &  $-$&  $-$& 32.2 &7.492&  $-$ &  $-$&  $-$ &  $-$&  $-$&  $-$\\
6137.002 &Fe I & 2.20 &$-$2.87 &cn&  $-$ &$-$  &  $-$ &$-$  &   $-$ & $-$ &  $-$&  $-$&  $-$ &  $-$&  $-$ &  $-$& 77.0 &7.304&  $-$&  $-$\\
6151.623 &Fe I & 2.18 &$-$3.28 &cn& 56.6 &7.213&  $-$ &$-$  &   $-$ & $-$ &  $-$&  $-$&  $-$ &  $-$&  $-$ &  $-$&  $-$ &  $-$&  $-$&  $-$\\
6157.733 &Fe I & 4.07 &$-$1.26 &nd&  $-$ &$-$  &  $-$ &$-$  &   $-$ & $-$ &  $-$&  $-$&  $-$ &  $-$&  $-$ &  $-$& 62.4 &7.334&  $-$&  $-$\\
6159.370 &Fe I & 4.61 &$-$1.97 &nd&  $-$ &$-$  &  $-$ &$-$  &  37.2 &7.494& 28.7&7.288&  $-$ &  $-$& 42.0 &7.643&  $-$ &  $-$&  $-$&  $-$\\
6165.363 &Fe I & 4.14 &$-$1.47 &cn& 42.5 &7.187& 98.5 &7.682&  82.3 &7.180&  $-$&  $-$& 80.5 &7.259& 71.6 &7.086&  $-$ &  $-$& 82.9&7.331\\
6173.341 &Fe I & 2.22 &$-$2.88 &cn& 79.3 &7.301&  $-$ &$-$  &   $-$ & $-$ &  $-$&  $-$&  $-$ &  $-$&  $-$ &  $-$&  $-$ &  $-$&  $-$&  $-$\\
6180.209 &Fe I & 2.73 &$-$2.58 &cn& 68.4 &7.327&  $-$ &$-$  &   $-$ & $-$ &  $-$&  $-$&  $-$ &  $-$&  $-$ &  $-$& 47.4 &7.013&  $-$&  $-$\\
6187.995 &Fe I & 3.94 &$-$1.72 &cn& 55.8 &7.482& 97.8 &7.657&  84.8 &7.208& 75.1&7.066& 70.1 &7.074& 83.5 &7.294&  $-$ &  $-$& 86.8&7.401\\
6200.321 &Fe I & 2.61 &$-$2.44 &cn& 74.4 &7.172&  $-$ & $-$ &   $-$ & $-$ &  $-$&  $-$&  $-$ &  $-$&  $-$ &  $-$& 64.5 &7.059&  $-$&  $-$\\
6213.437 &Fe I & 2.22 &$-$2.58 &nd& 89.2 &7.197&  $-$ & $-$ &   $-$ & $-$ &  $-$&  $-$&  $-$ &  $-$&  $-$ &  $-$&  $-$ &  $-$&  $-$&  $-$\\
6215.149 &Fe I & 4.19 &$-$1.13 &lz& 76.9 &7.551&  $-$ & $-$ &   $-$ & $-$ & 91.9&7.117&  $-$ &  $-$&  $-$ &  $-$& 44.8 &7.002&  $-$&  $-$\\
6229.232 &Fe I & 2.84 &$-$2.81 &cn& 42.7 &7.168&  $-$ & $-$ &   $-$ & $-$ &  $-$&  $-$& 86.5 &7.065&  $-$ &  $-$&  $-$ &  $-$& 99.2&7.313\\
6232.648 &Fe I & 3.65 &$-$1.22 &cn& 91.0 &7.365&  $-$ & $-$ &   $-$ & $-$ &  $-$&  $-$&  $-$ &  $-$&  $-$ &  $-$& 77.0 &7.154&  $-$&  $-$\\
6240.653 &Fe I & 2.22 &$-$3.27 &cn& 70.9 &7.514&  $-$ & $-$ &   $-$ & $-$ &  $-$&  $-$&  $-$ &  $-$&  $-$ &  $-$& 43.9 &7.099&  $-$&  $-$\\
6246.327 &Fe I & 3.60 &$-$0.88 &cn&  $-$ &$-$  &  $-$ & $-$ &   $-$ & $-$ &  $-$&  $-$&  $-$ &  $-$&  $-$ &  $-$& 94.2 &7.104&  $-$&  $-$\\
6270.231 &Fe I & 2.86 &$-$2.61 &cn& 63.1 &7.377&  $-$ & $-$ &   $-$ & $-$ &  $-$&  $-$&  $-$ &  $-$&  $-$ &  $-$& 50.1 &7.220&  $-$&  $-$\\
6301.508 &Fe I & 3.65 &$-$0.72 &cn&  $-$ &$-$  &  $-$ & $-$ &   $-$ & $-$ &  $-$&  $-$&  $-$ &  $-$&  $-$ &  $-$& 95.5 &6.972&  $-$&  $-$\\
6330.852 &Fe I & 4.73 &$-$1.74 &nd&  $-$ &$-$  &  $-$ & $-$ &   $-$ & $-$ &  $-$&  $-$&  $-$ &  $-$&  49.1&7.683&  $-$ &  $-$&  $-$&  $-$\\
6336.830 &Fe I & 3.69 &$-$0.86 &cn&  $-$ &$-$  &  $-$ & $-$ &   $-$ & $-$ &  $-$&  $-$&  $-$ &  $-$&  $-$ &  $-$& 85.8 &6.969&  $-$&  $-$\\
6344.155 &Fe I & 2.43 &$-$2.90 &cn&  $-$ &$-$  &  $-$ & $-$ &   $-$ & $-$ &  $-$&  $-$&  $-$ &  $-$&  $-$ &  $-$& 68.2 &7.379&  $-$&  $-$\\
6358.687 &Fe I & 0.86 &$-$4.17 &cn& 90.6 &7.299&  $-$ & $-$ &   $-$ & $-$ &  $-$&  $-$&  $-$ &  $-$&  $-$ &  $-$&  $-$ &  $-$&  $-$&  $-$\\
6380.750 &Fe I & 4.19 &$-$1.29 &cn& 59.8 &7.375&  $-$ & $-$ &   $-$ & $-$ &  $-$&  $-$& 97.5 &7.431& 85.2 &7.195& 39.2 &7.036&  $-$&  $-$\\
6408.026 &Fe I & 3.69 &$-$1.01 &cn&  $-$ &$-$  &  $-$ & $-$ &   $-$ & $-$ &  $-$&  $-$&  $-$ &  $-$&  $-$ &  $-$& 88.4 &7.165&  $-$&  $-$\\
6419.956 &Fe I & 4.73 &$-$0.24 &nd&  $-$ &$-$  &  $-$ & $-$ &   $-$ & $-$ &  $-$&  $-$&  $-$ &  $-$&  $-$ &  $-$& 74.0 &7.147&  $-$&  $-$\\
6481.878 &Fe I & 2.28 &$-$2.97 &cn& 78.4 &7.401&  $-$ & $-$ &   $-$ & $-$ &  $-$&  $-$&  $-$ &  $-$&  $-$ &  $-$& 59.1 &7.119&  $-$&  $-$\\
6551.676 &Fe I & 0.99 &$-$5.79 &nd&  $-$ &$-$  &  $-$ & $-$ &   $-$ & $-$ &  $-$&  $-$& 58.2 &7.187&  $-$ &  $-$&  $-$ &  $-$&  $-$&  $-$\\
6593.884 &Fe I & 2.43 &$-$2.42 &cn& 97.7 &7.381&  $-$ & $-$ &   $-$ & $-$ &  $-$&  $-$&  $-$ &  $-$&  $-$ &  $-$&  $-$ &  $-$&  $-$&  $-$\\
6597.561 &Fe I & 4.80 &$-$1.06 &lz& 40.4 &7.379&  $-$ & $-$ &  57.0 &7.171& 58.7&7.155& 60.3 &7.276& 51.2 &7.118& 29.6 &7.210& 66.1&7.401\\
6608.024 &Fe I & 2.28 &$-$4.04 &nd& 20.6 &7.306& 92.2 &7.638&   $-$ & $-$ &  $-$&  $-$&  $-$ &  $-$&  $-$ &  $-$&  $-$ &  $-$&  $-$&  $-$\\
6609.118 &Fe I & 2.56 &$-$2.66 &cn& 71.3 &7.233&  $-$ & $-$ &   $-$ & $-$ &  $-$&  $-$&  $-$ &  $-$&  $-$ &  $-$&  $-$ &  $-$&  $-$&  $-$\\
6646.932 &Fe I & 2.61 &$-$3.99 &nd& 14.0 &7.397&  $-$ & $-$ &   $-$ & $-$ &  $-$&  $-$& 48.4 &7.302& 50.3 &7.288&  $-$ &  $-$&  $-$&  $-$\\
6703.576 &Fe I & 2.76 &$-$3.16 &lz& 42.5 &7.400&  $-$ & $-$ &   $-$ & $-$ & 74.5&6.942& 95.8 &7.433& 92.6 &7.340&  $-$ &  $-$&  $-$&  $-$\\
6716.220 &Fe I & 4.58 &$-$1.93 &nd& 16.9 &7.467&  $-$ & $-$ &   $-$ & $-$ &  $-$&  $-$&  $-$ &  $-$&  $-$ &  $-$&  $-$ &  $-$&  $-$&  $-$\\
6725.353 &Fe I & 4.19 &$-$2.30 &nd& 15.0 &7.378& 55.9 &7.741&  42.2 &7.367& 26.8&7.039& 37.0 &7.362&  $-$ &  $-$&  $-$ &  $-$&  $-$&  $-$\\
6726.673 &Fe I & 4.61 &$-$1.00 &cn& 43.8 &7.186& 81.7 &7.441&  62.6 &6.958&  $-$& $-$ & 70.5 &7.155& 59.0 &6.954& 35.8 &7.084& 69.9&7.170\\
6733.151 &Fe I & 4.64 &$-$1.58 &lz& 18.4 &7.222& 64.8 &7.745&  52.1 &7.395& 32.7&6.991&  $-$ &  $-$& 52.5 &7.455&  $-$ &  $-$& 52.7&7.473\\
6745.090 &Fe I & 4.58 &$-$2.17 &nd&  $-$ &$-$  & 40.4 &7.812&   $-$ & $-$ & 19.9&7.211& 22.8 &7.398&  $-$ &  $-$&  $-$ &  $-$&  $-$&  $-$\\
6746.932 &Fe I & 2.61 &$-$4.25 &nd&  $-$ &$-$  &  $-$ & $-$ &   $-$ & $-$ &  $-$&  $-$& 23.0 &7.079&  $-$ &  $-$&  $-$ &  $-$&  $-$&  $-$\\
6752.716 &Fe I & 4.64 &$-$1.20 &bk& 29.1 &7.121&  $-$ & $-$ &   $-$ & $-$ &  $-$&  $-$&  $-$ &  $-$& 84.0 &7.625&  $-$ &  $-$&  $-$&  $-$\\
6786.856 &Fe I & 4.19 &$-$2.06 &nd& 26.6 &7.459& 71.6 &7.779&  47.6 &7.221&  $-$&  $-$& 55.3 &7.438& 46.7 &7.281&  $-$ &  $-$&  $-$&  $-$\\
6806.856 &Fe I & 2.73 &$-$3.21 &lz& 43.0 &7.421&  $-$ & $-$ &   $-$ & $-$ & 80.9&7.054& 92.6 &7.379& 89.8 &7.294& 25.7 &7.161& 99.7&7.515\\
6810.267 &Fe I & 4.61 &$-$0.99 &cn& 51.5 &7.311& 91.6 &7.605&  76.5 &7.177&  $-$&  $-$& 77.0 &7.249& 67.1 &7.077& 34.5 &7.040& 79.2&7.320\\
6828.596 &Fe I & 4.64 &$-$0.92 &lz& 59.0 &7.412&  $-$ & $-$ &  97.6 &7.503& 73.2&7.065& 91.4 &7.468& 93.8 &7.510& 45.2 &7.207&  $-$&  $-$\\
6839.835 &Fe I & 2.56 &$-$3.45 &lz& 42.9 &7.477&  $-$ & $-$ &   $-$ & $-$ & 76.6&6.997& 97.6 &7.483&  $-$ &  $-$&  $-$ &  $-$&  $-$&  $-$\\
6841.341 &Fe I & 4.61 &$-$0.75 &nd&  $-$ &$-$  &  $-$ & $-$ &   $-$ & $-$ &  $-$& $-$ &  $-$ &  $-$&  $-$ &  $-$& 61.6 &7.301&  $-$&  $-$\\
6842.689 &Fe I & 4.64 &$-$1.32 &lz&  $-$ &$-$  & 85.0 &7.853&  70.0 &7.438& 55.7&7.151& 67.4 &7.453& 67.6 &7.455&  $-$ &  $-$& 69.1&7.507\\
6843.655 &Fe I & 4.55 &$-$0.93 &lz& 62.3 &7.390&  $-$ & $-$ &  97.5 &7.394&  $-$&  $-$&  $-$ &  $-$& 88.0 &7.305& 49.8 &7.211& $-$ &  $-$\\
6858.155 &Fe I & 4.61 &$-$0.93 &cn& 59.7 &7.403& 97.6 &7.656&  83.6 &7.238&  $-$&  $-$& 81.5 &7.268& 85.5 &7.336&  $-$ &  $-$& 86.1&7.387\\
6999.885 &Fe I & 4.10 &$-$1.56 &nd&  $-$ &$-$  &  $-$ & $-$ &  91.9 &7.331&  $-$&  $-$& 90.7 &7.418&  $-$ &  $-$&  $-$ &  $-$& 96.1&7.549\\
7022.957 &Fe I & 4.19 &$-$1.25 &nd&  $-$ &$-$  &  $-$ & $-$ &  87.6 &7.066& 92.7&7.171&  $-$ &  $-$& 98.7 &7.344&  $-$ &  $-$&  $-$&  $-$\\
7071.866 &Fe I & 4.61 &$-$1.70 &lz&  $-$ &$-$  & 68.8 &7.891&  51.7 &7.463& 42.2&7.246&  $-$ &  $-$&  $-$ &  $-$&  $-$ &  $-$& 55.9&7.605\\
7112.170 &Fe I & 2.99 &$-$2.99 &cn&  $-$ &$-$  &  $-$ & $-$ &   $-$ & $-$ &  $-$&  $-$& 88.5 &7.400& 98.3 &7.532&  $-$ &  $-$&  $-$&  $-$\\
7132.985 &Fe I & 4.07 &$-$1.63 &cn& 47.5 &7.308& 93.8 &7.573&  75.3 &7.077&  $-$&  $-$& 69.6 &7.079& 77.6 &7.199& 35.5 &7.142& 77.6&7.234\\
7219.680 &Fe I & 4.07 &$-$1.35 &ow&  $-$ &$-$  &  $-$ & $-$ &   $-$ & $-$ & 93.6&7.119&  $-$ &  $-$&  $-$ &  $-$&  $-$ &  $-$&  $-$&  $-$\\
7284.842 &Fe I & 4.14 &$-$1.75 &nd&  $-$ &$-$  & 80.1 &7.528&   $-$ & $-$ & 70.8&7.193&  $-$ &  $-$&  $-$ &  $-$&  $-$ &  $-$&  $-$&  $-$\\
7306.570 &Fe I & 4.18 &$-$1.74 &lz&  $-$ &$-$  & 99.5 &7.909&  65.8 &7.168&  $-$& $-$ & 77.9 &7.454& 68.5 &7.285&  $-$ &  $-$& 72.9&7.388\\
7401.691 &Fe I & 4.19 &$-$1.60 &cn& 46.4 &7.366& 91.9 &7.638&  75.3 &7.189& 64.9&6.994& 76.1 &7.289& 73.2 &7.229& 32.9 &7.168& 92.7&7.604\\
7418.672 &Fe I & 4.14 &$-$1.38 &ow&  $-$ &$-$  &  $-$ & $-$ &  80.7 &6.987&  $-$&  $-$& 83.4 &7.126& 80.3 &7.061& 38.4 &7.001& 87.6&7.226\\
7443.026 &Fe I & 4.19 &$-$1.82 &nd&  $-$ &$-$  & 85.5 &7.742&   $-$ & $-$ &  $-$&  $-$&  $-$ &  $-$&  $-$ &  $-$& 25.8 &7.232&  $-$&  $-$\\
7583.796 &Fe I & 3.02 &$-$1.88 &cn& 89.2 &7.197&  $-$ & $-$ &   $-$ & $-$ &  $-$&  $-$&  $-$ &  $-$&  $-$ &  $-$&  $-$ &  $-$&  $-$&  $-$\\
7710.367 &Fe I & 4.22 &$-$1.11 &cn&  $-$ &$-$  &  $-$ & $-$ &  99.0 &7.107& 93.1&7.023&  $-$ &  $-$&  $-$ &  $-$& 54.1 &7.086&  $-$&  $-$\\
7723.210 &Fe I & 2.28 &$-$3.62 &fm& 39.9 &7.234&  $-$ & $-$ &   $-$ & $-$ &  $-$&  $-$& 98.3 &7.206&  $-$ &  $-$&  $-$ &  $-$&  $-$&  $-$\\
7746.605 &Fe I & 5.06 &$-$1.34 &nd&  $-$ &$-$  & 58.1 &7.874&   $-$ & $-$ &  $-$&  $-$& 34.0 &7.355& 42.8 &7.519&  $-$ &  $-$&  $-$&  $-$\\
7751.116 &Fe I & 4.99 &$-$0.72 &cn& 40.4 &7.175&  $-$ & $-$ &   $-$ & $-$ &  $-$&  $-$& 79.7 &7.440&  $-$ &  $-$& 33.8 &7.092&  $-$&  $-$\\
7780.568 &Fe I & 4.47 &$-$0.09 &cn&  $-$ &$-$  &  $-$ & $-$ &   $-$ & $-$ &  $-$&  $-$&  $-$ &  $-$&  $-$ &  $-$& 98.9 &7.062&  $-$&  $-$\\
7941.096 &Fe I & 3.27 &$-$2.58 &nd& 51.6 &7.445&  $-$ & $-$ &   $-$ & $-$ &  $-$&  $-$& 72.8 &7.036& 73.6 &7.005&  $-$ &  $-$&  $-$&  $-$\\
5991.378 &Fe II& 3.15 &$-$3.56 &cn& 37.6 &7.279& 46.8 &7.732&   $-$ & $-$ &  $-$&  $-$&  $-$ &  $-$&  $-$ &  $-$& 43.8 &7.346&  $-$&  $-$\\
6084.110 &Fe II& 3.20 &$-$3.97 &nd& 20.3 &7.318&  $-$ & $-$ &   $-$ & $-$ &  $-$&  $-$&  $-$ &  $-$&  $-$ &  $-$&  $-$ &  $-$&  $-$&  $-$\\
6149.249 &Fe II& 3.89 &$-$2.72 &cn& 37.8 &7.189&  $-$ & $-$ &   $-$ & $-$ &  $-$&  $-$&  $-$ &  $-$&  $-$ &  $-$& 45.4 &7.281& 47.4&7.536\\
6247.562 &Fe II& 3.89 &$-$2.26 &lh& 68.5 &7.371&  $-$ & $-$ &  51.7 &7.319&  $-$&  $-$& 64.5 &7.351&  $-$ &  $-$&  $-$ &  $-$& 62.1&7.412\\
6369.462 &Fe II& 2.89 &$-$4.36 &nd& 29.5 &7.628&  $-$ & $-$ &   $-$ & $-$ &  $-$&  $-$&  $-$ &  $-$&  $-$ &  $-$&  $-$ &  $-$&  $-$&  $-$\\
6416.928 &Fe II& 3.89 &$-$2.74 &cn&  $-$ &$-$  &  $-$ & $-$ &   $-$ & $-$ & 50.6&7.109&  $-$ &  $-$&  $-$ &  $-$&  $-$ &  $-$&  $-$&  $-$\\
6432.683 &Fe II& 2.89 &$-$3.58 &cn& 52.4 &7.344& 58.9 &7.738&  46.1 &7.308&  $-$&  $-$& 57.4 &7.354&  $-$ &  $-$& 53.4 &7.273& 51.2&7.316\\
6456.391 &Fe II& 3.90 &$-$2.07 &cn& 76.1 &7.335&  $-$ & $-$ &   $-$ & $-$ &  $-$&  $-$& 77.1 &7.435&  $-$ &  $-$&  $-$ &  $-$&  $-$&  $-$\\
7711.731 &Fe II& 3.90 &$-$2.47 &cn& 47.6 &7.087&  $-$ & $-$ &   $-$ & $-$ & 66.4&7.161&  $-$ &  $-$&  $-$ &  $-$&  $-$ &  $-$& 59.5&7.570\\
6154.230 &Na I & 2.10 &$-$1.57 &cn& 28.7 &6.007&125.2 &6.688&  86.4 &5.993& 53.0&5.829&  $-$ &  $-$& 83.3 &6.145& 24.7 &5.987& 96.6&6.369\\
6160.753 &Na I & 2.10 &$-$1.23 &cn& 55.1 &6.102&141.6 &6.583& 112.1 &5.993& 81.3&5.890&101.7 &6.123&104.5 &6.090& 34.8 &5.843&117.1&6.323\\
5528.418 &Mg I & 4.34 &$-$0.78 &nd&  $-$ &$-$  &  $-$ & $-$ &   $-$ & $-$ &  $-$&  $-$&  $-$ &  $-$&  $-$ &  $-$&193.0 &7.286&  $-$&  $-$\\
5711.095 &Mg I & 4.34 &$-$1.82 &nd& 80.3 &7.126&162.8 &7.780& 151.0 &7.416&136.7&7.463& 82.6 &6.606&  $-$ &  $-$& 88.4 &7.288&146.3&7.555\\
7657.606 &Mg I & 5.11 &$-$1.19 &nd& 91.3 &7.265&  $-$ & $-$ &   $-$ & $-$ &104.6&7.103&  $-$ &  $-$&111.5 &7.148& 80.8 &7.176&153.1&7.722\\
6696.020 &Al I & 3.14 &$-$1.33 &lw& 43.3 &6.211&  $-$ & $-$ &   $-$ & $-$ &  $-$&  $-$&  $-$ &  $-$&  $-$ &  $-$&  $-$ &  $-$&  $-$&  $-$\\
6698.670 &Al I & 3.14 &$-$1.87 &cn& 19.7 &6.275&105.9 &6.987&  71.4 &6.374& 43.0&6.207& 50.8 &6.321& 64.4 &6.452& 13.3 &6.134& 56.0&6.352\\
7835.317 &Al I & 4.02 &$-$0.58 &cn& 42.5 &6.258&129.3 &7.032& 110.5 &6.653& 67.6&6.304&  $-$ &  $-$&  $-$ &  $-$& 23.3 &5.944&  $-$&  $-$\\
7836.130 &Al I & 4.02 &$-$0.40 &cn& 48.7 &6.169&111.8 &6.625&  88.4 &6.201& 67.9&6.128& 96.5 &6.488& 68.5 &6.055& 40.9 &6.092&  $-$&  $-$\\
5665.563 &Si I & 4.92 &$-$2.04 &cn& 37.0 &7.382&103.6 & $-$ &   $-$ & $-$ &  $-$&  $-$&  $-$ &  $-$&  $-$ &  $-$& 32.8 &7.330& 78.9&7.918\\
5690.433 &Si I & 4.93 &$-$1.87 &cn&  $-$ &$-$  & 79.3 &7.882&   $-$ & $-$ &  $-$&  $-$&  $-$ &  $-$&  $-$ &  $-$&  $-$ &  $-$&  $-$&  $-$\\
5701.108 &Si I & 4.93 &$-$2.05 &cn& 51.6 &7.655&  $-$ & $-$ &  67.1 &7.749&  $-$&  $-$& 57.0 &7.524& 43.9 &7.329& 31.7 &7.328& 66.7&7.729\\
5772.149 &Si I & 5.08 &$-$1.67 &cn& 47.9 &7.349&  $-$ & $-$ &   $-$ & $-$ &  $-$&  $-$&  $-$ &  $-$&  $-$ &  $-$& 46.0 &7.338& 88.1&7.882\\
5793.079 &Si I & 4.93 &$-$1.95 &cn& 53.7 &7.582& 88.1 &8.112&  76.9 &7.810&  $-$&  $-$& 62.1 &7.505& 71.7 &7.701& 43.7 &7.438& 71.4&7.706\\
5948.548 &Si I & 5.08 &$-$1.19 &cn& 84.7 &7.447&  $-$ & $-$ &   $-$ & $-$ & 84.6&7.127&  $-$ &  $-$&120.5 &7.866& 74.7 &7.304&  $-$&  $-$\\
6125.026 &Si I & 5.61 &$-$1.54 &lz& 42.5 &7.642&  $-$ & $-$ &   $-$ & $-$ &  $-$&  $-$&  $-$ &  $-$& 77.7 &8.177&  $-$ &  $-$&  $-$&  $-$\\
6142.494 &Si I & 5.62 &$-$1.48 &ns&  $-$ &$-$  & 58.0 &7.940&   $-$ & $-$ &  $-$&  $-$&  $-$ &  $-$&  $-$ &  $-$& 31.8 &7.415& 50.8&7.686\\
6145.020 &Si I & 5.62 &$-$1.43 &ns& 41.2 &7.519&  $-$ & $-$ &  55.3 &7.771& 41.3&7.253&  $-$ &  $-$& 29.7 &7.229& 32.1 &7.370& 53.5&7.685\\
7034.910 &Si I & 5.87 &$-$0.81 &cn& 51.7 &7.279& 83.7 &8.001&   $-$ & $-$ & 53.7&7.135&  $-$ &  $-$& 60.1 &7.463& 51.9 &7.287&  $-$&  $-$\\
7226.208 &Si I & 5.61 &$-$1.30 &nd& 35.1 &7.226&  $-$ & $-$ &   $-$ & $-$ & 49.5&7.235& 61.4 &7.617&  $-$ &  $-$& 32.4 &7.189& 68.9&7.801\\
7405.790 &Si I & 5.61 &$-$0.68 &cn& 84.2 &7.360&118.4 &8.070& 110.9 &7.855& 97.6&7.404&  $-$ &  $-$&119.4 &7.884& 82.5 &7.334&115.6&7.889\\
7415.958 &Si I & 5.61 &$-$0.71 &cn& 94.9 &7.549&  $-$ & $-$ &   $-$ & $-$ &  $-$&  $-$&  $-$ &  $-$&  $-$ &  $-$& 93.4 &7.524&  $-$&  $-$\\
7918.383 &Si I & 5.95 &$-$0.54 &cn& 79.9 &7.433&  $-$ & $-$ &   $-$ & $-$ &  $-$&  $-$&  $-$ &  $-$&  $-$ &  $-$& 65.8 &7.245&  $-$&  $-$\\
7932.351 &Si I & 5.96 &$-$0.35 &cn&106.1 &7.576& 98.0 &7.823&   $-$ & $-$ &134.4&7.982& 83.7 &7.411& 93.4 &7.574& 94.9 &7.435&  $-$&  $-$\\
5512.989 &Ca I & 2.93 &$-$0.53 &cn&  $-$ &$-$  &  $-$ & $-$ & 132.5 &5.968&135.4&6.507&107.3 &5.885&136.1 &6.242& 80.8 &6.172&  $-$&  $-$\\
5581.979 &Ca I & 2.52 &$-$0.67 &cn&111.3 &6.364&  $-$ & $-$ & 192.2 &6.336&153.5&6.440&147.9 &6.179&184.3 &6.481& 98.7 &6.222&159.5&6.358\\
5588.764 &Ca I & 2.52 &   0.06 &cn&  $-$ &$-$  &  $-$ & $-$ & 191.1 &5.592&197.6&6.219&183.7 &5.855&186.5 &5.769&  $-$ &  $-$&190.1&5.953\\
5590.126 &Ca I & 2.52 &$-$0.70 &cn&102.6 &6.267&  $-$ & $-$ &   $-$ & $-$ &118.7&5.864&128.0 &5.908&133.2 &5.890& 93.1 &6.164&149.8&6.255\\
5601.286 &Ca I & 2.52 &$-$0.52 &cn&125.8 &6.411&  $-$ & $-$ & 162.9 &5.867&120.8&5.724&  $-$ &  $-$&161.5 &6.095&106.7 &6.191&  $-$&  $-$\\
5867.572 &Ca I & 2.93 &$-$1.61 &cn&  $-$ &$-$  & 81.9 &6.367&   $-$ & $-$ &  $-$&  $-$& 45.4 &5.942& 43.3 &5.830& 18.5 &6.101&  $-$&  $-$\\
6102.727 &Ca I & 1.88 &$-$0.79 &nd&138.6 &6.156&  $-$ & $-$ &   $-$ & $-$ &189.2&6.077&191.3 &5.886&  $-$ &  $-$&  $-$ &  $-$&  $-$&  $-$\\
6122.226 &Ca I & 1.89 &$-$0.32 &nd&185.2 &6.107&  $-$ & $-$ &   $-$ & $-$ &199.5&5.727&  $-$ &  $-$&  $-$ &  $-$&  $-$ &  $-$&  $-$&  $-$\\
6161.295 &Ca I & 2.52 &$-$1.19 &cn& 75.0 &6.264&  $-$ & $-$ &   $-$ & $-$ &129.5&6.452&112.4 &6.060&128.6 &6.221& 54.4 &5.981&  $-$&  $-$\\
6163.754 &Ca I & 2.52 &$-$1.07 &nd&  $-$ &$-$  &  $-$ & $-$ &   $-$ & $-$ &136.0&6.429&  $-$ &  $-$&  $-$ &  $-$&  $-$ &  $-$&  $-$&  $-$\\
6166.440 &Ca I & 2.52 &$-$1.19 &cn& 68.2 &6.144&130.4 &6.214& 133.1 &6.009&103.8&5.987&110.8 &6.030&121.9 &6.112& 57.3 &6.030&117.3&6.116\\
6169.044 &Ca I & 2.52 &$-$0.80 &lz& 97.9 &6.251&  $-$ & $-$ &   $-$ & $-$ &125.2&5.980&145.3 &6.168&125.2 &5.773& 82.0 &6.048&153.7&6.293\\
6169.564 &Ca I & 2.52 &$-$0.51 &cn&  $-$ &$-$  &  $-$ & $-$ &   $-$ & $-$ &128.4&5.750&163.3 &6.115&150.8 &5.853& 99.5 &6.048&158.5&6.071\\
6439.083 &Ca I & 2.52 &   0.16 &cn&168.9 &6.201&  $-$ & $-$ & 185.7 &5.328&166.5&5.625&193.5 &5.763&188.4 &5.604&148.9 &5.998&  $-$&  $-$\\
6449.820 &Ca I & 2.52 &$-$0.50 &cn&122.2 &6.316&  $-$ & $-$ &   $-$ & $-$ &  $-$&  $-$&  $-$ &  $-$&157.7 &5.915&  $-$ &  $-$&  $-$&  $-$\\
6455.605 &Ca I & 2.52 &$-$1.29 &cn&  $-$ &$-$  &  $-$ & $-$ & 110.9 &5.736&  $-$&  $-$& 93.3 &5.821&100.7 &5.847&  $-$ &  $-$&105.4&5.981\\
6471.668 &Ca I & 2.52 &$-$0.69 &cn& 94.9 &6.159&  $-$ & $-$ &   $-$ & $-$ &123.5&5.822&139.9 &6.006&  $-$ &  $-$& 83.7 &6.009&  $-$&  $-$\\
6493.788 &Ca I & 2.52 &$-$0.09 &cn&  $-$ &$-$  &  $-$ & $-$ &   $-$ & $-$ &168.1&5.952&178.6 &5.954&185.6 &5.942&116.9 &5.981&  $-$&  $-$\\
6499.654 &Ca I & 2.52 &$-$0.81 &cn& 88.6 &6.156&  $-$ & $-$ & 141.4 &5.750&106.0&5.620&127.5 &5.921&129.3 &5.856& 71.0 &5.893&139.2&6.096\\
6717.687 &Ca I & 2.71 &$-$0.52 &cn&  $-$ &$-$  &  $-$ & $-$ & 196.9 &6.333&156.8&6.364&  $-$ &  $-$&179.2 &6.393&107.0 &6.322&  $-$&  $-$\\
7148.150 &Ca I & 2.71 &$-$0.14 &cn&  $-$ &$-$  &  $-$ & $-$ &   $-$ & $-$ &  $-$&  $-$&191.7 &6.180&190.5 &6.058&137.2 &6.292&189.5&6.179\\
5526.821 &Sc II& 1.77 &   0.13 &nd&  $-$ &$-$  &122.6 &3.138& 124.5 &2.867&146.1&3.158&124.8 &3.011&142.4 &3.307& 96.5 &2.936&124.3&3.072\\
5657.880 &Sc II& 1.51 &$-$0.50 &nd& 81.5 &3.056&121.3 &3.384& 102.2 &2.734&104.2&2.617& 96.0 &2.766&102.4 &2.887& 85.1 &3.069&110.2&3.080\\
5684.198 &Sc II& 1.51 &$-$0.20 &nd&  $-$ & $-$ &  $-$ & $-$ &   $-$ & $-$ &  $-$&  $-$&  $-$ &  $-$&110.0 &2.725& 79.8 &2.668&  $-$&  $-$\\
6245.620 &Sc II& 1.51 &$-$0.98 &lz&  $-$ & $-$ &  $-$ & $-$ &   $-$ & $-$ & 97.2&2.894&  $-$ &  $-$& 98.0 &3.233&  $-$ &  $-$&  $-$&  $-$\\
6604.600 &Sc II& 1.36 &$-$1.16 &lz& 51.3 &2.926&114.9 &3.585& 102.3 &3.101&  $-$&  $-$& 86.2 &2.990& 90.8 &3.068& 51.1 &2.887&  $-$&  $-$\\
5866.461 &Ti I & 1.07 &$-$0.84 &cn& 42.2 &4.519&178.3 &5.360& 141.3 &4.315& 86.0&3.949&117.1 &4.444&110.6 &4.196& 36.0 &4.522&  $-$&  $-$\\
5953.170 &Ti I & 1.89 &$-$0.21 &cn& 50.6 &4.899&140.0 &5.158& 100.3 &4.140& 83.5&4.353& $-$  &  $-$&  $-$ &  $-$& 25.8 &4.515&104.1&4.623\\
6126.224 &Ti I & 1.07 &$-$1.32 &cn& 29.3 &4.731&129.7 &4.875& 131.7 &4.556&108.9&4.760& 93.7 &4.494&114.3 &4.689&  $-$ &  $-$&112.8&4.736\\
6258.110 &Ti I & 1.44 &$-$0.43 &cn& 57.8 &4.757&  $-$ & $-$ &   $-$ & $-$ &109.7&4.381&  $-$ &  $-$&  $-$ &  $-$& 40.2 &4.557&135.0&4.743\\
6261.106 &Ti I & 1.43 &$-$0.48 &cn& 65.1 &4.925&  $-$ & $-$ & 137.8 &4.313&109.2&4.406&121.3 &4.577&  $-$ &  $-$&  $-$ &  $-$&  $-$&  $-$\\
6312.241 &Ti I & 1.46 &$-$1.55 &lz&  $-$ &$-$  & 75.4 &4.693&  52.0 &4.147& 32.2&4.280& 40.0 &4.432&  $-$ &  $-$&  $-$ &  $-$& 57.5&4.597\\
6599.110 &Ti I & 0.90 &$-$2.08 &lz&  $-$ &$-$  &  $-$ & $-$ &  73.4 &4.146& 39.6&4.178& 62.5 &4.529&  $-$ &  $-$&  $-$ &  $-$& 84.9&4.736\\
6743.120 &Ti I & 0.90 &$-$1.63 &lz& 12.1 &4.332&133.9 &4.857&   $-$ & $-$ & 55.9&3.952& 84.1 &4.349&  $-$ &  $-$&  $-$ &  $-$&107.6&4.603\\
5727.057 &V  I & 1.08 &$-$0.01 &cn&  $-$ &$-$  &164.1 &4.431&   $-$ & $-$ &  $-$&  $-$&128.3 &3.907&194.5 &4.807& 41.1 &3.806&  $-$&  $-$\\
6090.216 &V  I & 1.08 &$-$0.14 &cn& 36.7 &3.710&138.6 &3.948& 109.7 &3.081&  $-$&  $-$& 91.6 &3.326& 81.6 &3.036& 18.3 &3.410&113.8&3.631\\
6216.358 &V  I & 0.28 &$-$0.75 &lz&  $-$ &$-$  &162.5 &3.772&   $-$ & $-$ &  $-$&  $-$&103.2 &2.940& 70.3 &2.349&  $-$ &  $-$&137.5&3.440\\
5783.866 &Cr I & 3.32 &$-$0.20 &cn& 45.4 &5.478&116.7 &5.902& 111.7 &5.570&  $-$&  $-$& 86.9 &5.451&106.3 &5.713& 32.6 &5.299&108.1&5.826\\
5787.926 &Cr I & 3.32 &$-$0.18 &cn& 38.5 &5.333&109.6 &5.749&  86.3 &5.118& 75.2&5.263& 70.7 &5.158& 62.1 &4.939& 31.7 &5.266& 87.8&5.430\\
6925.280 &Cr I & 3.45 &$-$0.33 &nd& 40.2 &5.587&  $-$ & $-$ &  61.7 &4.979& 60.9&5.246& 62.9 &5.269& 45.9 &4.930& 24.3 &5.322&  $-$&  $-$\\
6978.383 &Cr I & 3.46 &   0.14 &cn& 63.1 &5.536&  $-$ & $-$ &   $-$ & $-$ &  $-$&  $-$&  $-$ &  $-$&  $-$ &  $-$&  $-$ &  $-$&  $-$&  $-$\\
6979.806 &Cr I & 3.46 &$-$0.41 &lz& 30.6 &5.484&106.3 &5.952&   $-$ & $-$ & 73.1&5.538& 80.5 &5.637& 85.4 &5.655& 22.7 &5.371& 97.1&5.886\\
7355.891 &Cr I & 2.89 &$-$0.29 &cn& 63.0 &5.351&155.6 &5.806&   $-$ & $-$ & 89.9&4.927&  $-$ &  $-$&  $-$ &  $-$& 54.4 &5.272&152.4&5.860\\
7400.188 &Cr I & 2.90 &$-$0.17 &cn& 67.2 &5.315&164.7 &5.825& 137.9 &5.168&  $-$& $-$ &126.4 &5.369&120.9 &5.191&  $-$ &  $-$&140.6&5.571\\
6013.497 &Mn I & 3.07 &$-$0.15 &lz& 94.8 &5.413&177.0 &5.800& 140.0 &5.004&124.5&5.102&138.4 &5.270&136.7 &5.164& 67.2 &4.989&147.9&5.449\\
6021.803 &Mn I & 3.07 &   0.02 &lz& 68.5 &4.778&170.2 &5.555& 138.4 &4.810&121.0&4.864&137.7 &5.090&132.5 &4.933& 64.5 &4.773&147.3&5.271\\
5578.729 &Ni I & 1.68 &$-$2.80 &cn&  $-$ &$-$  &  $-$ & $-$ &   $-$ & $-$ &135.5&6.412&  $-$ &  $-$&146.8 &6.608&  $-$ &  $-$&124.7&6.324\\
5587.868 &Ni I & 1.93 &$-$2.14 &fm&  $-$ &$-$  &  $-$ & $-$ &   $-$ & $-$ &128.4&5.930&137.3 &6.167&  $-$ &  $-$& 62.5 &5.750&  $-$&  $-$\\
5593.746 &Ni I & 3.90 &$-$0.84 &cn& 51.9 &6.202& 93.2 &6.575&  77.1 &6.079& 66.4&5.835& 67.7 &5.988& 82.9 &6.254& 38.0 &5.990&  $-$&  $-$\\
5625.328 &Ni I & 4.09 &$-$0.70 &fm&  $-$ &$-$  &  $-$ & $-$ &   $-$ & $-$ &  $-$&  $-$&  $-$ &  $-$& 86.5 &6.400& 39.1 &6.056&  $-$&  $-$\\
5694.991 &Ni I & 4.09 &$-$0.61 &cn& 48.4 &6.091&  $-$ & $-$ &  85.1 &6.222& 64.9&5.802& 73.7 &6.085& 71.5 &6.046& 40.6 &5.991&  $-$&  $-$\\
5754.666 &Ni I & 1.93 &$-$2.33 &fm&  $-$ &$-$  &  $-$ & $-$ &   $-$ & $-$ &138.3&6.278&  $-$ &  $-$&130.0 &6.142& 69.5 &6.056&  $-$&  $-$\\
5805.226 &Ni I & 4.17 &$-$0.64 &cn&  $-$ &$-$  & 72.2 &6.297&  49.7 &5.742&  $-$&  $-$& 63.9 &6.033&  $-$ &  $-$&  $-$ &  $-$&  $-$&  $-$\\
6086.288 &Ni I & 4.26 &$-$0.53 &cn& 44.3 &6.088&  $-$ & $-$ &  62.5 &5.961& 73.0&6.060& 60.6 &5.963&  $-$ &  $-$& 29.8 &5.850& 67.2&6.116\\
6108.125 &Ni I & 1.68 &$-$2.63 &lz& 72.1 &6.033&149.6 &6.512& 147.3 &6.103&  $-$&  $-$&132.2 &6.138&  $-$ &  $-$& 53.9 &5.776&134.2&6.222\\
6111.078 &Ni I & 4.09 &$-$0.81 &fm& 30.6 &5.927& 76.5 &6.436&   $-$ & $-$ &  $-$&  $-$&62.9  &6.076&  $-$ &  $-$&  $-$ &  $-$& 74.3&6.316\\
6128.984 &Ni I & 1.68 &$-$3.33 &fm& 36.7 &6.074&117.8 &6.636&  89.0 &5.812& 80.1&5.718& 80.2 &5.903& 71.9 &5.725& 16.8 &5.699& 92.7&6.129\\
6130.141 &Ni I & 4.26 &$-$0.96 &cn& 25.6 &6.139& 63.3 &6.555&   $-$ & $-$ & 43.9&5.951& 38.8 &6.004&  $-$ &  $-$& 14.4 &5.872& 40.3&6.047\\
6176.816 &Ni I & 4.09 &$-$0.26 &cn&  $-$ &$-$  &  $-$ & $-$ &   $-$ & $-$ &  $-$&  $-$& 97.0 &6.117&  $-$ &  $-$& 52.9 &5.842& 99.6&6.228\\
6327.604 &Ni I & 1.68 &$-$3.11 &cn& 40.5 &5.911&132.5 &6.654& 120.0 &6.098& 94.6&5.737& 94.9 &5.915& 84.6 &5.699& 25.2 &5.686&108.7&6.184\\
6482.809 &Ni I & 1.93 &$-$2.63 &fm& 46.0 &5.785&132.6 &6.477& 110.8 &5.785&  $-$&  $-$&  $-$ &  $-$&  $-$ &  $-$&  $-$ &  $-$&  $-$&  $-$\\
6586.319 &Ni I & 1.95 &$-$2.73 &cn& 55.4 &6.071&120.0 &6.359&  90.3 &5.560&  $-$&  $-$&105.3 &6.039&  $-$ &  $-$& 27.4 &5.624&106.4&6.082\\
6635.150 &Ni I & 4.42 &$-$0.83 &lz& 25.7 &6.143& 86.2 &7.014&   $-$ & $-$ & 52.0&6.140&  $-$ &  $-$&  $-$ &  $-$& 20.9 &6.067&  $-$&  $-$\\
6643.638 &Ni I & 1.68 &$-$2.30 &fm&107.5 &6.300&  $-$ & $-$ & 168.1 &5.964&133.0&5.601&160.3 &6.131&  $-$ &  $-$& 87.4 &6.007&177.0&6.410\\
6767.784 &Ni I & 1.83 &$-$2.17 &cn& 82.3 &5.859&171.6 &6.421& 141.4 &5.638&  $-$&  $-$&  $-$ &  $-$&  $-$ &  $-$& 71.4 &5.732&148.8&6.061\\
6772.321 &Ni I & 3.66 &$-$0.95 &lz& 48.2 &5.938&  $-$ & $-$ &  87.6 &6.009& 63.9&5.528& 83.2 &6.014& 78.6 &5.937& 35.5 &5.753& 89.7&6.171\\
7001.600 &Ni I & 1.94 &$-$3.66 &nd& 13.4 &6.031& 85.6 &6.611&   $-$ & $-$ & 38.7&5.619& 48.0 &5.981& 43.5 &5.872&  $-$ &  $-$&  $-$&  $-$\\
7030.010 &Ni I & 3.54 &$-$1.73 &nd& 20.2 &5.995&  $-$ & $-$ &   $-$ & $-$ & 30.4&5.542& 39.4 &5.893&  $-$ &  $-$& 14.9 &5.891&  $-$&  $-$\\
7110.905 &Ni I & 1.93 &$-$2.92 &nd& 46.5 &6.036&147.6 &6.889& 142.2 &6.480&  $-$&  $-$&  $-$ &  $-$&125.7 &6.428& 20.6 &5.589&  $-$&  $-$\\
7122.206 &Ni I & 3.54 &   0.04 &cn&  $-$ &$-$  &  $-$ & $-$ &   $-$ & $-$ &162.5&6.031&  $-$ &  $-$&  $-$ &  $-$& 98.6 &5.691&194.5&6.417\\
7385.244 &Ni I & 2.74 &$-$1.97 &cn& 57.7 &6.128&145.6 &6.934& 108.7 &6.095& 74.0&5.509& 83.5 &5.840& 99.2 &6.078& 28.1 &5.646& 97.5&6.100\\
7414.514 &Ni I & 1.99 &$-$2.57 &cn&  $-$ &$-$  &  $-$ & $-$ &   $-$ & $-$ &120.3&5.923&  $-$ & $-$ &  $-$ &  $-$& 61.4 &6.073&  $-$&  $-$\\
7422.286 &Ni I & 3.63 &$-$0.33 &cn& 88.9 &5.939&180.4 &6.816& 154.5 &6.266&135.2&6.066&  $-$ & $-$ &151.4 &6.301& 82.4 &5.865&  $-$&  $-$\\
7525.118 &Ni I & 3.63 &$-$0.65 &cn& 78.8 &6.092&152.3 &6.823& 131.4 &6.284&120.8&6.143&  $-$ & $-$ &128.6 &6.323& 59.0 &5.796&  $-$&  $-$\\
7574.048 &Ni I & 3.83 &$-$0.61 &cn& 63.5 &5.986&116.3 &6.499&  93.9 &5.937& 84.6&5.709&101.9 &6.128& 93.0 &5.983& 51.2 &5.818&106.2&6.257\\
7714.310 &Ni I & 1.93 &$-$1.91 &cn&110.0 &6.085&  $-$ & $-$ &   $-$ & $-$ &178.1&6.094&  $-$ &  $-$&195.7 &6.251& 87.6 &5.779&  $-$&  $-$\\
7715.591 &Ni I & 3.70 &$-$0.95 &cn& 43.0 &5.832&123.8 &6.792&  68.3 &5.701& 66.0&5.567& 93.1 &6.164& 68.9 &5.770& 33.5 &5.704& 92.1&6.208\\
7727.616 &Ni I & 3.68 &$-$0.17 &cn&  $-$ &$-$  &154.5 &6.421&   $-$ & $-$ &  $-$&  $-$&130.3 &5.930&113.4 &5.664& 77.9 &5.663&  $-$&  $-$\\
7748.894 &Ni I & 3.70 &$-$0.33 &cn& 82.6 &5.883&141.1 &6.445& 121.5 &5.891&108.4&5.683&128.6 &6.086&101.2 &5.657& 74.7 &5.788&124.7&6.121\\
7788.933 &Ni I & 1.95 &$-$2.42 &fm&  $-$ &$-$  &196.8 &7.008& 155.9 &6.114&125.4&5.762&  $-$ &  $-$&137.1 &6.038& 75.3 &6.089&  $-$&  $-$\\
7797.588 &Ni I & 3.90 &$-$0.30 &lz& 78.0 &5.975&  $-$ & $-$ &  90.8 &5.656& 92.3&5.599&  $-$ &  $-$& 91.5 &5.722& 60.6 &5.724&  $-$&  $-$\\
6435.000 &Y  I & 0.07 &$-$0.82 &hl& 10.4 &2.560&126.9 &2.896& 132.1 &2.552& 96.0&2.300& 83.3 &2.354&126.0 &2.871&  $-$ &  $-$&108.2&2.642\\
6127.460 &Zr I & 0.15 &$-$1.06 &bg& 6.5  &2.833&119.6 &3.231& 110.8 &2.677&  $-$&  $-$&  $-$ &  $-$& 87.1 &2.769&  7.5 &3.042& 80.0&3.181\\
6134.570 &Zr I & 0.00 &$-$1.28 &bg& 4.4  &2.711&111.7 &3.031& 106.9 &2.568& 76.6&2.480& 82.3 &2.776&  $-$ &  $-$&  5.9 &2.997& 71.6&2.976\\
6140.460 &Zr I & 0.52 &$-$1.41 &bg&  $-$ &$-$  & 76.8 &3.444&  88.4 &3.307&  $-$&  $-$& 58.3 &3.435&  $-$ &  $-$&  $-$ &  $-$&  $-$&  $-$\\
6143.180 &Zr I & 0.07 &$-$1.10 &bg& 12.5 &3.097&125.3 &3.231& 151.5 &3.279&117.6&3.127&100.9 &2.980&132.5 &3.396& 15.6 &3.360& 92.2&3.292\\
5853.688 &Ba II& 0.60 &$-$1.01 &wm&112.5 &2.559&171.7 &2.888& 257.0 &3.110&268.5&3.215&182.6 &2.790&252.5 &3.214&160.0 &3.039&197.1&3.010\\
6141.727 &Ba II& 0.70 &$-$0.08 &wm&186.9 &2.423&249.9 &2.619& 477.9 &2.972&416.2&2.863&303.2 &2.639&485.0 &3.094&258.2 &2.745&311.0&2.755\\
6496.908 &Ba II& 0.60 &$-$0.38 &wm&178.7 &2.517&270.8 &2.854& 431.1 &3.013&378.4&2.921&288.9 &2.725&445.7 &3.164&222.3 &2.735&302.8&2.871\\
6390.480 &La II& 0.32 &$-$1.45 &lb&  $-$ &$-$  & 73.5 &1.999& 108.6 &2.292& 95.0&2.040& 68.2 &1.859& 88.1 &2.185& 32.9 &1.933& 81.4&2.109\\
6645.110 &Eu II& 1.37 &   0.20 &bc& 19.1 &0.776& 59.6 &1.236&  50.3 &0.829& 58.8&0.774& 27.2 &0.500& 45.6 &0.880& 16.1 &0.656& 48.2&0.938\\

\enddata
\tablenotetext{bc}{Bi$\acute{e}$mont et al. 1982}
\tablenotetext{bg}{Bi$\acute{e}$mont et al. 1981}
\tablenotetext{bk}{Bard \& Kock 1994}
\tablenotetext{cn}{Chen et al. 2000b}
\tablenotetext{fm}{Fuhr et al. 1988}
\tablenotetext{hl}{Hannaford et al. 1982}
\tablenotetext{lb}{Luck \& Bond 1991}
\tablenotetext{lh}{Lambert et al. 1996}
\tablenotetext{lw}{Lambert \& Warner 1968}
\tablenotetext{lz}{Liang et al. 2003}
\tablenotetext{nd}{NIST database (http://physics.nist.gov)}
\tablenotetext{ns}{Nissen \& Schuster 1997}
\tablenotetext{ow}{O'Brian et al. 1991}
\tablenotetext{wm}{Weise \& Martin 1980}
\end{deluxetable}
\end{document}